\documentclass[prd,preprint,nofootinbib,superscriptaddress]{revtex4}

\usepackage{amsmath,amssymb}
\usepackage{graphicx}
\usepackage{dcolumn}
\usepackage{bm}
\usepackage{cancel}

\begin{document}
\title{Neutrino Masses from an $A_4$ Symmetry\\ in Holographic
Composite Higgs Models}

\date{\today}

\preprint{CAFPE-128/10} 
\preprint{UGFT-258/10} 

\author{Francisco del \'Aguila}
\author{Adri\'an Carmona}
\affiliation{
CAFPE and Departamento de F\'{\i}sica Te\'orica y del 
Cosmos, \\
Universidad de Granada, E-18071 Granada, Spain}
\author{Jos\'e Santiago}
\affiliation{
CAFPE and Departamento de F\'{\i}sica Te\'orica y del 
Cosmos, \\
Universidad de Granada, E-18071 Granada, Spain}
\affiliation{Institute for Theoretical Physics, ETH, CH-8093, 
Z\"urich, Switzerland}

\begin{abstract}
We show that
holographic composite Higgs Models with a discrete $A_4$ symmetry  
naturally predict hierarchical charged lepton masses and an
approximate tri-bimaximal
lepton mixing with the correct scale of neutrino masses. 
They also satisfy current constraints from electroweak precision
tests, lepton flavor violation and lepton mixing in a large 
region of parameter space. 
Two phenomenologically relevant features 
arise in
these models. 
First, 
an extra suppression on the lepton Yukawa couplings makes 
the $\tau$ lepton more composite than naively expected from its mass.  
As a consequence new light \textit{leptonic} resonances, 
with masses as low as few hundreds of GeV, large
couplings to $\tau$ and a very characteristic collider phenomenology,
are quite likely.
Second,
the discrete symmetry $A_4$ together with the model structure
provide a double-layer of flavor protection that allows to keep 
tree-level mediated processes below present experimental limits. 
One-loop processes violating lepton flavor, 
like $\mu \to e \gamma$, 
may be however observable at future experiments.
\end{abstract}


\maketitle

\section{Introduction}

The primary goal of the Large Hadron Collider (LHC) is the study of
the precise mechanism of electroweak symmetry breaking (EWSB). One
interesting possibility is that new gauge interactions become
strongly coupled at the TeV scale, breaking some global symmetries but
not necessarily the electroweak gauge symmetry. The Higgs boson can
then arise as a composite Goldstone boson of the spontaneous global
symmetry breaking. The coupling to an
elementary sector (external to the strongly coupled theory) breaks
explicitly the global symmetries and
generates a potential for the Higgs at the loop
level~\cite{Kaplan:1983fs}. The AdS/CFT
correspondence~\cite{Maldacena:1997re} suggests that models with
warped extra dimensions~\cite{Randall:1999vf}
are weakly coupled duals
to strongly coupled four-dimensional (4D) conformal
theories~\cite{ArkaniHamed:2000ds}, and therefore they provide a
calculable framework for composite Higgs
models~\cite{Contino:2003ve}.~\footnote{ 
See~\cite{compositeH:4D} for a 
discussion of composite Higgs models from the effective 
4D point of view.}
A recent review of tools employed to study models with warped extra
dimensions and their phenomenological implications can be found 
in~\cite{Davoudiasl:2009cd}.

Using a custodially symmetric set-up~\cite{Agashe:2003zs} 
and a fermionic content that guarantees protection of the 
$Z\bar{b}_L b_L$ coupling~\cite{Agashe:2006at}
(see~\cite{Djouadi:2006rk}
for an alternative), minimal
composite Higgs models from warped extra
dimensions~\cite{Agashe:2004rs} have been shown to
dynamically generate EWSB at the expense of a 
modest fine-tuning~\cite{Agashe:2004rs,Medina:2007hz,Csaki:2008zd}.
Furthermore, they are fully compatible with electroweak precision
tests (EWPT)~\cite{Agashe:2005dk,Carena:2006bn} and can even be easily
extended to accommodate dark
matter~\cite{Agashe:2007jb,Panico:2008bx,Carena:2009yt}.
(How relevant is the fine-tuning is a debatable matter as it has been
shown that there is a large intersection in these models 
among the regions with a good
pattern of EWSB, the correct top mass and a good behaviour under
EWPT~\cite{Medina:2007hz,Panico:2008bx}.)
Most of the studies related to the five-dimensional (5D) 
realization of composite Higgs models have only focused on the quark
sector. In fact, although some of the first studies of bulk fermion 
phenomenology in models with warped extra dimensions
were made with the leptonic sector in
mind~\cite{Grossman:1999ra,Huber:2000ie,Huber:2003sf}, not much
progress has been made until quite recently. In particular, older 
proposals for models of lepton masses have, with few exceptions, 
not been updated to make them compatible with new,
realistic models in warped extra dimensions. 
One possible reason is that
the generation of Yukawa couplings by fermion 
splitting~\cite{ArkaniHamed:1999dc} seemed 
to naturally lead to a hierarchical pattern of fermion masses and
mixing angles, like the one observed in the quark sector, but not to large
mixing angles 
like those 
observed in the neutrino sector. 
This was recently shown not to be
necessarily true~\cite{Agashe:2008fe}, and a realization of neutrino masses
within this framework and with a realistic dark matter
candidate was presented in~\cite{Carena:2009yt}.

An alternative approach to differentiate the quark and lepton 
spectra is to assume a global symmetry acting on the leptonic
sector. 
4D models of neutrino masses with an $A_4$ 
symmetry~\cite{A4in4D,LFV:A4} 
can predict a tri-bimaximal (TBM)~\cite{Harrison:2002er} 
pattern of lepton mixing to leading
order (LO), what agrees quite well with
observation~\cite{GonzalezGarcia:2007ib,Schwetz:2008er}.
This global symmetry can be also implemented in simple models with warped
extra dimensions~\cite{Csaki:2008qq}.~\footnote{Other symmetries that 
can simultaneously accommodate the pattern of quark and lepton 
mixing have been also considered in 5D 
contexts~\cite{Feruglio:2007uu,Chen:2009gy} and in models compatible
with an underlying GUT structure~\cite{deMedeirosVarzielas:2005ax}.} 
Such a construction presents an
advantage over 4D models since the mass hierarchy 
follows from wave-function overlapping, geometrically realizing 
the required
Frogatt-Nielsen mass generation in 4D models. 
Besides, it also improves other 5D models 
that solely rely on the former for it has
an extra built-in flavor protection due to the discrete $A_4$ symmetry. 
Our goal is to extend this set-up to models of gauge-Higgs
unification (GHU), which are arguably the most natural models of EWSB 
in warped extra dimensions. We will show that, despite
some subtleties related to the way fermions acquire
non-trivial Yukawa couplings in GHU models~\cite{Csaki:2008zd}, it
is easy to find examples that naturally generate a realistic fermion 
spectrum also in the lepton sector.

Two new features phenomenologically relevant 
come out from our analysis. First, due to
an extra suppression of the leptonic Yukawa couplings implied by
the $A_4$ symmetry, the $\tau$ lepton is typically more composite than
one would naively expect from its mass. This makes new 
leptonic resonances at the electroweak scale a likely occurrence in these
models. Besides, as they come in two almost degenerate doublets with 
hypercharges $-1/2$ and $-3/2$, respectively, and mainly couple to 
$\tau$, they provide a very distintive signature at LHC 
for they only decay through definite channels and into $\tau$ leptons. 
This structure is dictated by the same symmetry that protects the 
$Z \bar{b}_L b_L$
coupling in this type of models~\cite{Agashe:2006at}, 
which in the leptonic sector protects the Standard Model (SM) 
lepton couplings despite the large new lepton couplings to 
$\tau$~\cite{Atre:2008iu}. 
Second, the $A_4$ symmetry together with the protecting 
mechanism above~\cite{Agashe:2009tu,Buras:2009ka} 
result in a double-layer flavor protection. 
Thus, lepton flavor violation (LFV) 
mediated by tree-level exchange of heavy modes is 
further suppressed, and typically below current experimental limits. 
The main constraints result from one-loop
processes, like $\mu \to e \gamma$, which is close but quite often
below the present experimental sensitivity, being then within the
reach of future experiments.

The outline of the paper is as follows. We describe the model in
Section~\ref{model}. The leptonic spectrum is computed in
Section~\ref{leptonic:spectrum}, where the LO implications 
of the $A_4$ symmetry are discussed in detail. The 
corrections to these LO results are classified in
Section~\ref{higher:orders}; and the constraints from EWPT 
and flavor observables are considered in Section~\ref{constraints}, 
where we also give an explicit example of a realistic model. 
Section~\ref{conclusions} is left to our conclusions, and 
some technical details are collected in the appendices.   

\section{The model \label{model}}

We consider a 5D model in a slice of $AdS_5$ with metric
\begin{equation}
\mathrm{d}s^2
=
a^2(z)
(
\eta_{\mu\nu}\mathrm{d}x^\mu\mathrm{d}x^\nu
-\mathrm{d}z^2)
=\left(\frac{R}{z}\right)^2
(
\eta_{\mu\nu}\mathrm{d}x^\mu\mathrm{d}x^\nu
-\mathrm{d}z^2) 
\end{equation}
and $R \leq z \leq R^\prime$, where $R \sim M_{P}^{-1}$ and $R^\prime \sim
\mathrm{TeV}^{-1}$ are the location of the UV and IR brane,
respectively. 
Following~\cite{Agashe:2004rs}, we 
assume an $SO(5)\times U(1)_X$ bulk gauge symmetry broken by boundary
conditions to $SO(4)\times U(1)_X$ on the IR brane and to $SU(2)_L
\times U(1)_Y$ on the UV brane. These read
\begin{eqnarray}
\begin{array}{c}L_{\mu}^a(+,+),\qquad B_{\mu}(+,+),\\
R_{\mu}^b(-,+),\qquad Z^{\prime}_{\mu}(-,+),\end{array}\qquad
C^{\hat{a}}_{\mu}(-,-), 
\label{contor}
\end{eqnarray}
where $-$ ($+$) stands for Dirichlet (Neumann) boundary conditions at
the corresponding brane.  
The superscripts $a=1,2,3$, $b=1,2$ label the
$SO(4)$ gauge bosons in explicit $SU(2)_L \times SU(2)_R$ notation, 
and 
\begin{equation}
B_{\mu}=\frac{g_X R_{\mu}^3+g_5 X_{\mu}}{\sqrt{g_5^2+g_X^2}},\qquad
Z^{\prime}_{\mu}=\frac{g_5 R_{\mu}^3-g_X
  X_{\mu}}{\sqrt{g_5^2+g_X^2}}, 
\end{equation}
with $g_5$ and $g_X$ the 5D $SO(5)$ and $U(1)_X$ gauge couplings,
respectively. (The electric charge $Q=T^3_L + Y = T^3_L + T^3_R + Q_X$ 
with this normalization.)  
Finally, $C^{\hat{a}}_\mu$, 
$\hat{a}=1,\ldots,4$, 
are the gauge bosons corresponding to the
$SO(5)/SO(4)$ coset space. 

The gauge directions along $SO(5)/SO(4)$ are
broken on both branes and there is a
massless zero mode along the $5$-th component,
\begin{equation}
C_5^{\hat{a}}(x,z)=
\sqrt{\frac{2/R}{1-(R/R^\prime)^2}}
\frac{z}{R^{\prime}} h^{\hat{a}}(x)
+\ldots
\approx
\sqrt{\frac{2}{R}}\frac{z}{R^{\prime}}
h^{\hat{a}}(x)
+\ldots~,
\end{equation}
where the dots denote massive modes. 
(We have chosen the normalization
constant to obtain a canonically normalized scalar, and in the second
equality we have used $R \ll R^\prime$.) 
These four scalars transform
as a $\mathbf{4}$ of $SO(4)$ and are identified with the SM Higgs.
5D gauge invariance guarantees that any potential generated for these
scalars has to arise from non-local contributions and therefore, it is
finite to all orders in perturbation theory~\cite{vonGersdorff:2002rg}. 

Regarding the matter content of the model, there are several
possibilities. We consider here all fermions to be in fundamental
representations of $SO(5)$. 
Thus, four multiplets per family are required in order to have
independent localizations for left and right-handed zero
modes. This construction is parallel to the one giving rise to
realistic composite Higgs models in the quark 
sector~\cite{Contino:2006qr,Panico:2008bx}; 
and as we will show, 
a similar matter content transforming non-trivially under a global 
$A_4$ symmetry generates the observed leptonic spectrum in a natural way,  
without conflict with present experimental data. 
Hence, there are four 5D fermion representations per generation 
transforming as the fundamental $SO(5)$ representation {\bf 5}, with 
boundary conditions 
\begin{eqnarray}
\zeta_{1}&=&
\left(\begin{array}{cc} \widetilde{X}_{1}[-+]&  
\nu_1[++]\\ \tilde{\nu}_{1}[-+]& e_1[++]\end{array}
\right)
\oplus \nu^{\prime}_{1}[-+],
\quad
\zeta_{2}=
\left(\begin{array}{cc} 
\widetilde{X}_{2}[+-]&  \nu_2[+-]\\ 
\tilde{\nu}_{2}[+-]& e_2[+-]\end{array}
\right)
\oplus \nu^{\prime}_{2}[--],\nonumber\\
\zeta_{3}&=&
\left(\begin{array}{cc} 
\nu_{3}[-+] &\tilde{e}_{3}[-+]\\
e_{3}[-+]& \widetilde{Y}_{3}[-+]\end{array}
\right)
\oplus e^{\prime}_{3}[-+],
\quad \zeta_{\alpha}=
\left(\begin{array}{cc} 
\nu_{\alpha}[+-] &
\tilde{e}_{\alpha}[+-]\\
e_{\alpha}[+-]& 
\widetilde{Y}_{\alpha}[+-]\end{array}
\right)
\oplus e^{\prime}_{\alpha}[--], \label{multiplets}
\end{eqnarray}
where $\zeta_{1,2}$ and $\zeta_{3,\alpha}$ have $U(1)_X$ charge $0$
and $-1$, respectively. 
Note that there are three copies for each $\zeta_{1,2,3}$ 
because there are three 
families, but only one $\zeta_{\alpha}$ set with $\alpha$ running 
over the three lepton flavors $e,\mu,\tau$. 
We explicitly show the decomposition under $SU(2)_L\times
SU(2)_R$, {\bf 5} = ({\bf 2},{\bf 2}) $\oplus$ ({\bf 1},{\bf 1}). 
The bi-doublet is represented by a
$2 \times 2$ matrix
with the $SU(2)_L$ rotation acting vertically and the $SU(2)_R$ one 
horizontally (\textit{i.e.} 
the left and right columns correspond to fields with
$T^3_R=\pm 1/2$, whereas the upper and lower components
have $T^3_L=\pm 1/2$, respectively). The bi-doublets in $\zeta_{1,2}$
contain two $SU(2)_L$ doublets of hypercharge $\frac{1}{2}$ and
$-\frac{1}{2}$, and those in $\zeta_{3,\alpha}$
two $SU(2)_L$ doublets of hypercharge $-\frac{1}{2}$ and
$-\frac{3}{2}$, respectively. 
The corresponding electric charges read 
\begin{equation}
Q(\nu.)=
Q(\tilde{\nu}.)=Q(\nu.^\prime)=0,
\quad
Q(e.)=Q(e.^\prime)=Q(\tilde{e}.)=-1,
\quad
Q(\tilde{X}.)=+1,
\quad
Q(\tilde{Y}.)=-2,
\end{equation}
where the dot denotes all possible values of the corresponding subscript.
The signs in square brackets are
a shorthand for the boundary conditions.
A Dirichlet boundary condition for the right-handed (RH) component is
denoted by $[+]$, whereas $[-]$ denotes a Dirichlet boundary condition for 
the left-handed (LH) chirality. Finally, the
first sign corresponds to the boundary condition at the 
UV brane and the second one at the IR brane.
The chosen boundary conditions allow for a LH zero mode 
transforming as an $SU(2)_L$
doublet with hypercharge $-1/2$ in $\zeta_1$, 
a RH singlet of charge $-1$ in
$\zeta_\alpha$, and a RH neutral singlet in $\zeta_2$.

As generally in $A_4$ models, an extra global
symmetry must be imposed to forbid dangerous operators. 
A discrete $Z_8$ group does the job in our case. 
Both global symmetries will be
broken at the two branes by localized scalars transforming as 
gauge singlets, $\phi$ and $\eta$ at the
UV brane and $\phi^\prime$ and $\eta^\prime$ at the IR one. 
The fermion and scalar transformation properties 
under $A_4 \times Z_8$ are gathered in Table~\ref{cuatro}. 
\begin{table}[!h]
\begin{center}
\begin{tabular}{ccc}
& $A_4$&$Z_8$ 
\\
$\zeta_1$& $\mathbf{3}$&$1$
\\
$\zeta_2$& $\mathbf{3}$&$2$
\\
$\zeta_3$& $\mathbf{3}$&$1$
\\
$\zeta_{\alpha}$&
$\mathbf{1},\mathbf{1}^{\prime},\mathbf{1}^{\prime\prime}$&$4$
\end{tabular}
\hspace{1cm}
\begin{tabular}{ccc}
&$A_4$&$Z_8$\\
$\phi$(UV)&$\mathbf{3}$&4\\
$\eta$(UV)&$\mathbf{1}$&4\\
$\phi^{\prime}$(IR)&$\mathbf{3}$&5\\
$\eta^{\prime}$(IR)&$\mathbf{1}$&7
\end{tabular}
\caption{Bulk fermion (left) and localized scalar (right) quantum 
number assignments under the discrete group $A_4 \times Z_8$.}
\label{cuatro}
\end{center}
\end{table}
The three copies of $\zeta_{1,2,3}$ span the $A_4$ triplet 
representation, whereas each $\zeta_\alpha$ 
transforms as one the three different $A_4$ one-dimensional representations 
(see Appendix~\ref{a4} for a summary of the 
$A_4$ representations). 

Once the matter content is fixed, we can
write down the most general Lagrangian compatible with the
symmetries. 
The bulk Lagrangian reads 
\begin{equation}
\mathcal{L}=
\int_R^{R^{\prime}}
\mathrm{d}z
\,a^4 
\left\{
\overline{\zeta}_k
\left[
i \cancel{D} + \left( D_z+2\frac{a^\prime}{a}\right)\gamma^5
- a M_k
\right] \zeta_k
+
\overline{\zeta}_\alpha
\left[
i \cancel{D} + \left(  D_z+2\frac{a^\prime}{a}\right)\gamma^5
- a M_\alpha \right] \zeta_\alpha
\right\} ,
\end{equation} 
where summation on repeated indices 
$k\in\{1,2,3\}$, $\alpha\in\{e,\mu,\tau\}$ is understood. 
$D_{\mu,z}$ are the gauge
covariant derivatives and the bulk Dirac masses are conventionally 
parametrized in terms of the fundamental scale $R$, 
\begin{equation}
M_{k,\alpha}=\frac{c_{k,\alpha}}{R}.
\end{equation}
Note
that the $A_4$ symmetry implies a family independent bulk mass
for $\zeta_k$. The most general localized Lagrangians, excluding
kinetic terms (discussed below), compatible with
the boundary conditions, and local and global symmetries, 
can be written 
\begin{eqnarray}
-\mathcal{L}_{UV}&=&\frac{x_{\eta}}{2\Lambda}\eta
\overline{\nu}_{2R}^{\prime\,c} \nu_{2R}^{\prime}
+\frac{x_{\nu}}{2\Lambda}\phi 
\overline{\nu}_{2R}^{\prime\,c} \nu_{2R}^{\prime}
+x_l
\overline{l}_{1L}l_{3R}
+\mathrm{h.c.}+\ldots~,
\nonumber \\ 
-\mathcal{L}_{IR}&=&\left(\frac{R}{R^{\prime}}\right)^4
\left\{ \frac{y_{b}^{\alpha}}{\Lambda^{\prime}}
\left[
\left(\overline{l}_{3 L} \phi^{\prime}\right)^{\alpha} l_{\alpha R}
+
\left(\overline{\tilde{l}}_{3 L} \phi^{\prime}\right)^{\alpha} 
\tilde{l}_{\alpha R}
\right]+\frac{y_s^{\alpha}}{\Lambda^{\prime}}
\left(\overline{e}^\prime_{3 L} \phi^{\prime}\right)^{\alpha} e^\prime_{\alpha R}
\right.\nonumber\\
 &&
\phantom{\left(\frac{R}{R^{\prime}}\right)^4}
+ \left. \frac{y_b}{\Lambda^{\prime}}
\left[
\eta^{\prime\,}\overline{l}_{1L} l_{2R}
+\eta^{\prime\,}\overline{\tilde{l}}_{1L} \tilde{l}_{2R}
\right]
+\frac{y_{s}}{\Lambda^{\prime}}
\eta^{\prime\,}
\overline{\nu}^{\prime}_{1L} \nu^{\prime}_{2R} 
\right\}+\mathrm{h.c.}+\ldots~,
\end{eqnarray}
where we have assumed that lepton number is only violated on the UV
brane.~\footnote{This assumption, which corresponds to the strong 
sector preserving lepton number, can be obtained as an accidental 
symmetry by introducing, for instance, larger $SO(5)$ representations.}  
$l$ denotes the SM-like doublet and $\tilde{l}$ stands for 
the other $SU(2)_L$ doublet within the given bi-doublet,
whereas the dots correspond to higher dimensional operators. 
We have also introduced the LH and RH chirality projections 
$\zeta_{L,R}\equiv [(1\mp \gamma^5)/2] \zeta$, 
recovering the standard 4D notation. 
Finally, 
$\left(\phantom{A}\right)^{\alpha},~\alpha=e,\mu,\tau$, 
are the $\mathbf{3}\times \mathbf{3}$ combinations 
transforming under $A_4$ as 
$\mathbf{1},\mathbf{1}^{\prime\prime}$ and $\mathbf{1}^{\prime}$, 
respectively. 

As usually in these models, we shall assume that $A_4\times Z_8$ 
is spontaneously broken by the boundary scalar 
vacuum expectation values (v.e.v.) 
\begin{eqnarray}
\langle \phi\rangle=(v,0,0),
\qquad \langle \eta \rangle = v_{\eta}, 
\qquad \langle \phi^{\prime}\rangle=
(v^{\prime},v^{\prime},v^{\prime})
\qquad \textrm{and} 
\qquad\langle\eta^{\prime}\rangle=v_{\eta}^{\prime},
\end{eqnarray}
resulting in the brane localized terms 
\begin{eqnarray}
\label{UVIR}
-\mathcal{L}_{UV}&=&
\frac{1}{2}\overline{\nu}_{2R}^{\prime\,c}\theta_M
\nu_{2R}^{\prime}+x_l\bar{l}_{1L}l_{3R}
+\mathrm{h.c.}+\ldots~,
\nonumber \\
-\mathcal{L}_{IR}&=&\left(\frac{R}{R^{\prime}}\right)^4
\left\{
\sqrt{3} \frac{v^\prime}{\Lambda^\prime}
\left[
\overline{l}_{3L} \Omega
\begin{pmatrix} 
y_b^e & 0 & 0 \\
0&y_b^\mu &  0 \\
 0 & 0 &y_b^\tau 
\end{pmatrix}
l_R
+[l_3,l \to \tilde{l}_3,\tilde{l}]
+
\overline{e}_{3L}^{\prime}
\Omega 
\begin{pmatrix} 
y_s^e & 0 & 0 \\
0&y_s^\mu &  0 \\
 0 & 0 &y_s^\tau 
\end{pmatrix}
e^{\prime}_R
\right]
\right.\nonumber\\
&&
\phantom{\left(\frac{R}{R^{\prime}}\right)^4}
\left. +
y_b \frac{v_\eta^\prime}{\Lambda^\prime}
\big[
\overline{l}_{1L}
l_{2R}
+
\overline{\tilde{l}}_{1L}
\tilde{l}_{2R}
\big]
+
y_s \frac{v_\eta^\prime}{\Lambda^\prime}
\overline{\nu}^{\prime}_{1L}
\nu^{\prime}_{2R}
\right\}
+\mathrm{h.c.}+\ldots~, 
\end{eqnarray}
with the Majorana mass matrix 
\begin{eqnarray}
\theta_M\equiv\left(\begin{array}{ccc}\frac{x_{\eta}v_{\eta}}{\Lambda}&0&0\\
0&\frac{x_{\eta}v_{\eta}}{\Lambda}&\frac{x_{\nu}v}{\Lambda}\\
0&\frac{x_{\nu}v}{\Lambda}&\frac{x_{\eta}v_{\eta}}{\Lambda}
\end{array}\right)=\left(
\begin{array}{ccc}\epsilon_s&0&0\\0&\epsilon_s&
\epsilon_t\\0&\epsilon_t&\epsilon_s\end{array}\right), 
\end{eqnarray}
and the unitary matrix 
\begin{eqnarray}
\Omega\equiv\frac{1}{\sqrt{3}}
\left(\begin{array}{ccc}1&1&1\\1&\omega&\omega^2\\
1&\omega^2&\omega\end{array}\right),\qquad
\omega=e^{2\pi i/3}. 
\end{eqnarray}
In order to simplify Eq. (\ref{UVIR}), we can rotate the 
matter fields  
\begin{eqnarray}
\label{rotation}
\zeta_{k}\to \Omega~\zeta_k,\qquad \forall~k,
\end{eqnarray}
leaving the bulk lagrangian $\mathcal{L}$ invariant.
However, the localized terms  
\begin{eqnarray}
-\mathcal{L}_{UV}&=&
\frac{1}{2}\overline{\nu}_{2R}^{\prime\,c}\hat{\theta}_M
\nu_{2R}^{\prime}+x_l\bar{l}_{1L}l_{3R}
+\mathrm{h.c.}+\ldots , \nonumber \\
-\mathcal{L}_{IR}&=&\left(\frac{R}{R^{\prime}}\right)^4
\left[
\sqrt{3} \frac{v^\prime}{\Lambda^\prime}\big(
 y_b^\alpha 
\,\overline{l}_{3\alpha L}
l_{\alpha R}
+
y_b^\alpha 
\,
\overline{\tilde{l}}_{3\alpha L} 
\tilde{l}_{\alpha R}
+
y_s^\alpha
\,
\overline{e}_{3\alpha L}^{\prime}
e^{\prime}_{\alpha R}
\big)
\right.\nonumber\\
&&
\phantom{\left(\frac{R}{R^{\prime}}\right)^4}
\left. +
y_b \frac{v_\eta^\prime}{\Lambda^\prime}
\big(
\overline{l}_{1L}
l_{2R}
+
\overline{\tilde{l}}_{1L}
\tilde{l}_{2R}
\big)
+
y_s \frac{v_\eta^\prime}{\Lambda^\prime}
\overline{\nu}^{\prime}_{1L}
\nu^{\prime}_{2R}
\right]
+\mathrm{h.c.}+\ldots~, 
\end{eqnarray}
become diagonal in flavor space (the terms proportional to 
$x_l$ and $y_{b,s}$ are actually flavor independent), 
except for the Majorana masses 
\begin{eqnarray}
\label{Majterm}
\hat{\theta}_M\equiv \Omega\theta_M\Omega
=\left(\begin{array}{ccc}\epsilon_s+\frac{2\epsilon_t}{3}
&-\frac{\epsilon_t}{3}&-\frac{\epsilon_t}{3}\\ 
-\frac{\epsilon_t}{3}&\frac{2\epsilon_t}{3}
&\epsilon_s-\frac{\epsilon_t}{3}\\ 
-\frac{\epsilon_t}{3}&\epsilon_s-\frac{\epsilon_t}{3}
&\frac{2\epsilon_t}{3}\end{array}\right).
\end{eqnarray}
Dirichlet boundary conditions are modified in the presence of these
boundary terms. Thus, on the UV brane 
\begin{equation}
l_{1R}-x_l l_{3R}=0,\quad
l_{3L}+x_l^\ast l_{1L}=0,
\quad
\nu_{2L}^{\prime}+\hat{\theta}_M^\dagger \nu_{2R}^{\prime\,c}=0,
\label{majo}
\end{equation}
and on the IR one 
\begin{eqnarray}
l_{3\alpha R}
+
\sqrt{3}\frac{v^\prime}{\Lambda^\prime} y^{\alpha}_b
l_{\alpha R}&=&0,\quad
\tilde{l}_{3\alpha R}+
\sqrt{3}\frac{v^\prime}{\Lambda^\prime} y^{\alpha}_b
\tilde{l}_{\alpha R}=0,\quad
e_{3\alpha R}^{\prime}
+\sqrt{3}\frac{v^\prime}{\Lambda^\prime} y^{\alpha}_s
e^{\prime}_{\alpha R}=0,\nonumber\\ 
l_{\alpha L}-
\sqrt{3}\frac{v^\prime}{\Lambda^\prime} y^{\alpha\,\ast}_b
l_{3\alpha L}
&=&0,\quad \,\,\;
\tilde{l}_{\alpha L}-
\sqrt{3}\frac{v^\prime}{\Lambda^\prime} y^{\alpha\,\ast}_b
\tilde{l}_{3\alpha L}=0,
\quad \,\,
e^{\prime}_{\alpha L}-
\sqrt{3}\frac{v^\prime}{\Lambda^\prime} y^{\alpha\,\ast}_s
e_{3\alpha L}^{\prime}=0,\nonumber\\ 
l_{1R}+
y_b\frac{v_\eta^\prime}{\Lambda^\prime}
l_{2R}&=&0,
\qquad \quad 
\tilde{l}_{1R}
+y_b\frac{v_\eta^\prime}{\Lambda^\prime}
\tilde{l}_{2R}=0,
\qquad \quad
\nu_{1R}^{\prime}
+y_s\frac{v_\eta^\prime}{\Lambda^\prime}
\nu^{\prime}_{2R}=0,
\nonumber\\
l_{2L}-
y_b^\ast\frac{v_\eta^\prime}{\Lambda^\prime}
l_{1L}&=&0,
\qquad \quad 
\tilde{l}_{2L}-
y_b^\ast\frac{v_\eta^\prime}{\Lambda^\prime}
\tilde{l}_{1L}=0,
\quad \qquad
\nu^{\prime}_{2L}-
y_s^\ast\frac{v_\eta^\prime}{\Lambda^\prime}
\nu_{1L}^{\prime}=0.
\label{ir}
\end{eqnarray}
From these equations we observe that 
the lepton doublet zero mode is shared by all
multiplets due to the non-zero values of 
$x_l$, $y_b$ and $y^\alpha_b$, 
whereas $y_s$ splits the RH neutrino zero mode
between $\zeta_1$ and $\zeta_2$, 
and $y^\alpha_s$ splits the RH charge
$-1$ singlet between $\zeta_3$ and $\zeta_\alpha$. 
This splitting is
crucial in models of GHU, since the Higgs being part of a gauge
multiplet can only mix fermion fields within the same $SO(5)$
multiplet, coupling to them with the same (gauge) strength. 
The non-trivial flavor
structure is then only due to the brane terms above.
Thus, 
the only source of flavor violation in the rotated basis comes from
$\hat{\theta}_M$ in Eq. (\ref{Majterm}), 
whose particular form will eventually lead to TBM 
mixing in the leptonic sector.   
This flavor universality is a welcome consequence of the
$A_4$ symmetry, for it will also prevent flavor violating operators
mediated by heavy KK gauge bosons to exceed current experimental bounds. 
This observation, which was made in simpler models with warped extra
dimensions~\cite{Csaki:2008qq}, is maintained at this order in the more
realistic models with GHU under study here. 
Incidentally, the extra fields required to complete the $SO(5)$ 
representations imply that simpler $Z_2$ or $Z_3$ symmetries are not 
suitable to banish operators violating this mixing pattern. 

\section{The leptonic spectrum \label{leptonic:spectrum}}

In order to find the lepton masses and mixings we have to solve
the equations of motion derived from the bulk action with the
boundary conditions in Eqs. (\ref{majo}) and (\ref{ir}). This 
can be actually carried out exactly in the case of GHU models because 
the Higgs, which is part of a higher-dimensional gauge field, can be 
eliminated from the bulk by a rotation in gauge space, 
thus reducing the Higgs effect to the modification of the boundary conditions. 
This is essential, for otherwise the Higgs would mix different multiplets 
in the bulk, and the corresponding equations of motion would be 
forbiddingly difficult to solve. 
Still, the large number of fields involved makes the solution of the full
system technically challenging. An alternative approach is to
perform a Kaluza-Klein (KK) expansion without including the Higgs, 
and then to take into account its effects by diagonalizing the 
corresponding mass matrix. 
In this case one must include the Majorana masses not in the KK expansion 
but as a contribution to the mass matrix. 
Otherwise we would have to incorporate the effect of all 
physical modes up to the Majorana mass scale (which is $\sim R^{-1}$) 
in order to obtain an accurate enough approximation~\cite{Huber:2003sf}. 
Furthermore, the leading contribution to the \textit{light} neutrino masses
and mixing angles can be obtained by simply considering the zero
modes in the KK expansion 
(thus including the {\it heavy} Majorana RH neutrinos), 
for heavier KK modes give a suppressed contribution. 
This so called zero mode approximation (ZMA) is
convenient because of the transparent way the flavor structure leading to
TBM mixing is realized. We will thus proceed in three steps,
first we will compute the light lepton masses and mixing angles in the
ZMA. Then, we will include the  massive KK modes 
but still incorporating the localized Majorana masses and the Higgs effects 
in the mass matrix. 
Finally, we will take these into account considering the boundary conditions 
directly in the KK expansion.

The Yukawa couplings, being originally gauge couplings, are flavor diagonal 
and do not mix different 5D multiplets
\begin{eqnarray}
\mathcal{L}_Y&=&
g_5 h^{\hat{a}}(x)
\sqrt{\frac{2}{R}}\frac{1}{\sqrt{1-(R/R^\prime)^2}}
\int_R^{R^{\prime}}\mathrm{d}z\left(\frac{R}{z}\right)^4\frac{z}{R^\prime}
\left(\bar{\zeta}_kT_C^{\hat{a}}\Gamma^5\zeta_k
+\bar{\zeta}_{\alpha}T_C^{\hat{a}}\Gamma^5\zeta_{\alpha}\right)
\nonumber\\ 
&=& 
-ig_5 v_H \sqrt{\frac{2}{R}}\frac{1}{\sqrt{1-(R/R^\prime)^2}}
\int_R^{R^{\prime}}\mathrm{d}z
\left(\frac{R}{z}\right)^4\frac{z}{R^\prime}
\left(\bar{\zeta}_{kL}T_C^{4}\zeta_{kR}
+\bar{\zeta}_{\alpha L}T_C^{4}\zeta_{\alpha R}\right)+\mathrm{h.c.}~,
\end{eqnarray}
where we have used in the last equality $\Gamma^5= -i \gamma^5$, and 
assumed that the Higgs takes a v.e.v. 
$\langle h^{\hat{a}}\rangle =v_H \delta^{\hat{a},4}$.
Neglecting $R/R^\prime \ll 1$ and inserting the expression for 
$T_C^4$  in Appendix~\ref{SO5}, we get the Yukawa Lagrangian 
from the bulk 
\begin{eqnarray}
\label{yuk}
\mathcal{L}_Y&=&
\frac{g_5v_H}{2} \sqrt{\frac{2}{R}}
\int_R^{R^{\prime}}\mathrm{d}z\left(\frac{R}{z}\right)^4
\frac{z}{R^\prime}
\bigg\{
\sum_{s=1,2}
\Big[\overline{\nu}_{sL}^{\prime}
\left(\nu_{sR}+\tilde{\nu}_{sR}\right)
-\left(\overline{\nu}_{sL}
+\overline{\tilde{\nu}}_{sL}\right)\nu_{sR}^{\prime}\Big]
\\
&&
\phantom{\frac{g_5v_H}{2}
\sqrt{\frac{2}{R}}
\int_R^{R^{\prime}}
\mathrm{d}z\left(\frac{R}{z}\right)^4\frac{z}{R^\prime}
}
+\sum_{s=3,\alpha}
\Big[\overline{e}_{s L}^{\prime}\left(e_{s R}+
\tilde{e}_{s R}\right)
-\left(\overline{e}_{s L}
+\overline{\tilde{e}}_{s L}\right)e_{s R}^{\prime}\Big]
\bigg\}+\mathrm{h.c.}.
\nonumber
\end{eqnarray}

\subsection{Lepton spectrum in the ZMA}

In this section we only consider the leptonic zero modes. 
The localized Majorana masses and the Higgs couplings will be
incorporated as mass terms to be diagonalized. The localized Dirac masses,
on the other hand, have to be taken into account exactly. Since they
mix different multiplets through the boundary conditions, the physical
zero modes (the same will happen for massive modes) are split among
all multiplets mixed by them. 
In particular, the LH lepton doublets live in all four
multiplets. Note that as we do not include in the expansion the Majorana mass
term, which is the only source of flavor violation,  
different generations do not mix. 
The properly normalized zero modes satisfying the
corresponding boundary conditions read
\begin{eqnarray}
l_{1\alpha L}(x,z)&=&
\frac{1}{\sqrt{R^{\prime}}}
\left(\frac{z}{R}\right)^2
\left(\frac{z}{R^{\prime}}\right)^{-c_1}
\frac{f_{c_1}}{\sqrt{\iota_\alpha}}l_{\alpha L}(x)+\ldots~,
\label{l1L:zeromode}
\\ 
l_{2\alpha L}(x,z)&=&
y_b^\ast\frac{v_\eta^\prime}{\Lambda^{\prime}}
\frac{1}{\sqrt{R^{\prime}}}\left(\frac{z}{R}\right)^{2}
\left(\frac{z}{R^{\prime}}\right)^{-c_2}
\frac{f_{c_1}}{\sqrt{\iota_\alpha}}l_{\alpha L}(x)+\ldots~,
\label{l2L:zeromode}
\\
l_{3\alpha L}(x,z)&=&
-x_l^\ast\frac{1}{\sqrt{R^{\prime}}}
\left(\frac{R}{R^{\prime}}\right)^{c_3-c_1}
\left(\frac{z}{R}\right)^2
\left(\frac{z}{R^\prime}\right)^{-c_3}
\frac{f_{c_1}}{\sqrt{\iota_\alpha}}l_{\alpha L}(x)+\ldots~,
\label{l3L:zeromode}
\\
l_{\alpha L}(x,z)&=&
-\sqrt{3}x_l^\ast\frac{v^{\prime}}{\Lambda^{\prime}}
\frac{1}{\sqrt{R^{\prime}}}
\left(\frac{R}{R^{\prime}}\right)^{c_3-c_1}\left(\frac{z}{R}\right)^2
\left(\frac{z}{R^{\prime}}\right)^{-c_{\alpha}}
y_{b}^{\alpha\,\ast}\frac{f_{c_1}}{\sqrt{\iota_\alpha}}l_{\alpha L}(x)+\ldots~,
\label{laL:zeromode}
\end{eqnarray}
where $\alpha=e,\mu,\tau$ denote the lepton flavor and 
$l_{1,2,3}(x,z)$, 
$l_\alpha(x,z)$ stand for the doublet component of 
hypercharge $-1/2$ within each $\zeta_{1,2,3,\alpha}$,
respectively. 
Then, $l_\alpha(x)$ are the {\it physical} zero
modes; and the dots correspond to heavy KK modes. 
The flavor dependent term takes the form  
\begin{equation}
\iota_{\alpha}
\equiv
1+|y_b|^2\frac{v_\eta^{\prime\,2}}{\Lambda^{\prime 2}}
\frac{f_{c_1}^2}{f_{c_2}^2}
+|x_l|^2
\left(\frac{R}{R^{\prime}}\right)^{2(c_3-c_1)}
\left[\frac{f_{c_1}^2}{f_{c_3}^2}+|y_b^{\alpha}|^2
\frac{3v^{\prime 2}}{\Lambda^{\prime 2}}
\frac{f_{c_1}^2}{f_{c_{\alpha}}^2}\right], 
\end{equation}
with  
\begin{equation}
f_c \equiv \left[
  \frac{1-2c}{1-\left(\frac{R}{R^\prime}\right)^{1-2c}}
\right]^{\frac{1}{2}} 
\end{equation}
defined as usual. 
Eqs.~(\ref{l1L:zeromode}-\ref{laL:zeromode}) show that $x_l$ governs the
splitting of the LH lepton doublet zero mode between $\zeta_{1,2}$ and
$\zeta_{3,\alpha}$. Similarly, the splitting between $\zeta_1$ and
$\zeta_2$ and the one between $\zeta_3$ and $\zeta_\alpha$ are governed
by $y_b$ and $y_b^\alpha$, respectively. Also note that for $c_3 >
c_1 $ the zero mode components along $\zeta_{3,\alpha}$ have 
an extra suppression proportional to 
$(R/R^\prime)^{c_3-c_1}$. 

The RH charged lepton zero modes live in the $SO(4)$ singlet component
of $\zeta_3$ and $\zeta_\alpha$, 
\begin{eqnarray}
e^{\prime}_{3\alpha  R}(x,z)&=&  
-\sqrt{3}y_s^{\alpha}\frac{v^{\prime}}{\Lambda^{\prime}}
\frac{1}{\sqrt{R^{\prime}}}
\left(\frac{z}{R}\right)^2
\left(\frac{z}{R^{\prime}}\right)^{c_3}
\frac{f_{-c_{\alpha}}}
{\sqrt{
\rho_\alpha
}}
e_{\alpha R}(x) +
\ldots, \label{eR3:zeromode}
\\
e^{\prime}_{\alpha R}(x,z)&=& 
\frac{1}{\sqrt{R^{\prime}}}
\left(\frac{z}{R}\right)^2
\left(\frac{z}{R^{\prime}}\right)^{c_{\alpha}}
\frac{f_{-c_{\alpha}}
}{\sqrt{
\rho_\alpha
}}
e_{\alpha R}(x) +
\ldots~, \label{eRa:zeromode}
\end{eqnarray}
with
\begin{equation}
\rho_\alpha \equiv 1+|y_s^{\alpha}|^2\frac{3v^{\prime 2}}{\Lambda^{\prime 2}}
\frac{f_{-c_{\alpha}}^2}{f_{-c_3}^2}.
\end{equation}

Finally, there are RH neutrino zero modes living in the $SO(4)$
singlet components of $\zeta_{1,2}$, which 
read
\begin{eqnarray}
\nu^{\prime}_{1\alpha  R}(x,z)&=& 
-y_s \frac{v_\eta^\prime}{\Lambda^\prime}
\frac{1}{\sqrt{R^{\prime}}}
\left(\frac{z}{R}\right)^2\left(\frac{z}{R^{\prime}}\right)^{c_1}
\frac{f_{-c_2}}{\sqrt{
\lambda
}} 
\nu_{\alpha R}(x) + \ldots,
\\
\nu^{\prime}_{2\alpha  R}(x,z)&=& 
\frac{1}{\sqrt{R^{\prime}}}
\left(\frac{z}{R}\right)^2\left(\frac{z}{R^{\prime}}\right)^{c_2}
\frac{f_{-c_2}}{\sqrt{
\lambda
}} 
\nu_{\alpha R}(x) + \ldots~,
\end{eqnarray}
with
\begin{equation}
\lambda \equiv
1+|y_s|^2\frac{v_\eta^{\prime\,2}}{\Lambda^{\prime 2}}
\frac{f_{-c_2}^2}{f_{-c_1}^2}.
\end{equation}
Note that these profiles 
are not only flavor diagonal but flavor independent.
 
We can now insert the former expressions in the general
Yukawa Lagrangian, Eq. (\ref{yuk}), and get the 
corresponding zero mode mass term 
\begin{eqnarray}
-\mathcal{L}_Y=\bar{e}_L\mathcal{M}_D^e
e_R+\bar{\nu}_L
\mathcal{M}_D^{\nu}\nu_R+\mathrm{h.c.},
\end{eqnarray}
with 
\begin{eqnarray}
\left(\mathcal{M}_D^e\right)_{\alpha\beta}&=&
\frac{\sqrt{3}g_5 v_H x_l}{2\sqrt{2R}}
\frac{v^{\prime}}{\Lambda^{\prime}}
\left(\frac{R}{R^{\prime}}\right)^{c_3-c_1}
(y_{s}^{\alpha}-y_{b}^{\alpha})
\frac{f_{c_1}f_{-c_\alpha}}
{\sqrt{\iota_{\alpha}\rho_{\alpha}}}\delta_{\alpha\beta},
\label{MDe:ZMA}
\\
\left(\mathcal{M}_D^{\nu}\right)_{\alpha\beta}
&=&
-\frac{g_5 v_H v_\eta^\prime}{2\sqrt{2R}\Lambda^{\prime }}
\left(y_s-y_b\right)
\frac{f_{c_1}f_{-c_2}}{
\sqrt{\lambda \iota_{\alpha}}}\delta_{\alpha\beta}.
\label{MDnu:ZMA}
\end{eqnarray}
On the other hand, the UV brane term 
\begin{eqnarray}
-\mathcal{L}_M=
\left.\frac{1}{2}
\overline{\nu}_{2R}^{\prime\,c}
\hat{\theta}_M
\nu_{2R}^{\prime}\right|_{R}+\mathrm{h.c.}  
\label{UVMajorana:mass}
\end{eqnarray}
gives a Majorana mass contribution to the three RH neutrinos, 
so that the total zero mode mass Lagrangian writes 
\begin{equation}
\label{Majorana}
-\mathcal{L}_m=
\bar{e}_L\mathcal{M}_D^e e_R
+\bar{\nu}_L\mathcal{M}_D^{\nu}\nu_R
+\frac{1}{2}\overline{\nu}_R^c\mathcal{M}_M^{\nu}\nu_R
+\mathrm{h.c.},
\end{equation}
with 
\begin{eqnarray}
\mathcal{M}_M^{\nu}\equiv 
\frac{f_{-c_2}^2}{\lambda R^{\prime}}
\left(\frac{R}{R^{\prime}}\right)^{2c_2}\hat{\theta}_M.
\end{eqnarray}
Assuming $\lambda \approx 1$ and $R/R^\prime \approx 10^{-16}$, 
the Majorana mass scale is in the range
$(10^{-2}-10^{-5})/R \approx 10^{17}-10^{14}~\mbox{GeV}$ 
for $-0.5\leq c_2 \leq -0.35$.
$\mathcal{L}_m$ is already diagonal for charged leptons 
(see Eq. (\ref{MDe:ZMA})).
Furthermore, the localization parameters $f_{-c_\alpha}$
naturally explain a hierarchical pattern of charged lepton
masses.~\footnote{The $A_4$ symmetry forces the LH charged leptons to 
share a common localization thus naturally explaining why the 
mass hierarchy in this sector is smaller than the one in the 
charge $2/3$ quark sector~\cite{Santiago:2008vq}.}
The electron and muon masses are easily obtained with the
corresponding zero modes localized towards the UV brane. The
tau mass induces some tension that requires $c_1$ and $c_3$ to
be relatively close to $1/2$, $c_{1,3} \lesssim 0.6$, and the RH tau
to be somewhat localized towards the IR brane, $c_\tau \geq -1/2$. 
This tension is 
stronger the smaller the factor
$(y^{\alpha}_s-y^{\alpha}_b) x_l
v^\prime/\Lambda^\prime$ is. Note the 
$v^\prime/\Lambda^\prime$ suppression due to the $A_4$ structure.
This suppression makes the RH tau generically more composite ($c_\tau>
-1/2$) than naively expected from its mass. 
What generically implies {\it light} leptonic resonances accessible 
at the LHC, as discussed in section~\ref{constraints}. 
The $c_3-c_1$ difference also controls how the LH zero modes are
split between the $\zeta_{1,2}$ and $\zeta_{3,\alpha}$ multiplets
(see~\cite{Contino:2006qr} for a related discussion).
This becomes essential to protect the $\tau$ (LFV) couplings to the 
$Z$ when it is near the IR brane. 

Let us now turn our attention to the
neutrino sector. 
The matrix elements in Eq. (\ref{Majorana}) satisfy 
$\Vert\mathcal{M}_D^{\nu}\Vert 
\sim \mathcal{O}\left(\mathrm{TeV}\right) 
\ll \Vert \mathcal{M}_M^{\nu}\Vert\lesssim 
\mathcal{O}\left(M_{\textrm{Pl}}\right)$ 
for natural values of the model parameters. 
Then, integrating out the {\it heavy} RH neutrinos we 
obtain the standard see-saw type Majorana mass matrix 
for the LH neutrinos 
\begin{eqnarray}
\tilde{\mathcal{M}}^{\nu}&=&
-\mathcal{M}_D^{\nu}\left.
\mathcal{M}_M^{\nu}\right.^{-1}
\left(\mathcal{M}_D^{\nu}\right)^T
\nonumber\\ 
&=&
-\frac{\tilde{m}}{3}\left(
\begin{array}{rrr}
\frac{1}{\iota_e}\left[\frac{1}{\epsilon_s}
+ \frac{2}{\epsilon_s+\epsilon_t}\right] &
\frac{1}{\sqrt{\iota_e\iota_{\mu}}}
\left[\frac{1}{\epsilon_s}-\frac{1}{\epsilon_s+\epsilon_t}\right]
&
\frac{1}{\sqrt{\iota_e\iota_{\tau}}}
\left[\frac{1}{\epsilon_s}-\frac{1}{\epsilon_s+\epsilon_t} 
\right]\\ 
 \frac{1}{\sqrt{\iota_e
\iota_{\mu}}}
\left[\frac{1}{\epsilon_s}
-\frac{1}{\epsilon_s+\epsilon_t}\right]
 &\frac{1}{\iota_{\mu}}\left[
   \frac{1}{\epsilon_s}-\frac{1}{\epsilon_s-\epsilon_t}
-\frac{\epsilon_t}{\Delta}\right] 
 & 
 \frac{1}{\sqrt{\iota_{\mu}\iota_{\tau}}}\left[\frac{1}{\epsilon_s}
   +\frac{1}{\epsilon_s-\epsilon_t}
   +\frac{\epsilon_s}{\Delta}\right]\\ 
\frac{1}{\sqrt{\iota_e\iota_{\tau}}}
\left[\frac{1}{\epsilon_s}-\frac{1}{\epsilon_s+\epsilon_t}
\right]&
\frac{1}{\sqrt{\iota_{\mu}\iota_{\tau}}}\left[\frac{1}{\epsilon_s}
  +\frac{1}{\epsilon_s-\epsilon_t}  +
  \frac{\epsilon_s}{\Delta}\right]& 
\frac{1}{\iota_{\tau}}\left[
  \frac{1}{\epsilon_s}
-\frac{1}{\epsilon_s-\epsilon_t}-\frac{\epsilon_t}{\Delta}\right] 
\end{array}
\right),\nonumber\\
\end{eqnarray}
where we have defined $\Delta=\epsilon_s^2-\epsilon_t^2$ and
\begin{eqnarray}
\tilde{m}
\equiv
\frac{g_5^2}{R} \frac{(y_s-y_b)^2
  v_\eta^{\prime\,2}}{8\Lambda^{\prime\, 2}} v_H^2 R^\prime 
f_{c_1}^2
\left(\frac{R}{R^{\prime}}\right)^{-2c_2}
=
g^2 \log\left(\frac{R^\prime}{R}\right) \frac{(y_s-y_b)^2
  v_\eta^{\prime\,2}}{8\Lambda^{\prime\, 2}} v_H^2 R^\prime 
f_{c_1}^2
\left(\frac{R}{R^{\prime}}\right)^{-2c_2}.
\label{emetil}
\end{eqnarray}
In the last equality we have used the tree-level 
matching of the 5D and 4D
gauge coupling constants (in the absence of brane kinetic terms) 
\begin{eqnarray}
g_5=g\sqrt{R \log (R^{\prime}/R)}~, \label{matching}
\end{eqnarray}
with $g\approx 0.65$ the 4D $SU(2)_L$ coupling constant.
If we choose $c_{1,3}\geq \frac{1}{2}, c_3 > c_\alpha$, we can take  
$\iota_{\alpha}\cong \iota$ independent of $\alpha$ 
and then 
\begin{eqnarray}
\tilde{\mathcal{M}}^{\nu}\cong-\frac{\tilde{m}}{3\iota}\left(
\begin{array}{rrr}
\frac{1}{\epsilon_s}+ \frac{2}{\epsilon_s+\epsilon_t} &
\frac{1}{\epsilon_s}-\frac{1}{\epsilon_s+\epsilon_t}
&\frac{1}{\epsilon_s}-\frac{1}{\epsilon_s+\epsilon_t} \\ 
 \frac{1}{\epsilon_s}-\frac{1}{\epsilon_s+\epsilon_t} &
 \frac{1}{\epsilon_s}-\frac{1}{\epsilon_s-\epsilon_t}
-\frac{\epsilon_t}{\Delta}& 
 \frac{1}{\epsilon_s} +\frac{1}{\epsilon_s-\epsilon_t}
 +\frac{\epsilon_s}{\Delta}\\ 
\frac{1}{\epsilon_s}-\frac{1}{\epsilon_s+\epsilon_t} &
\frac{1}{\epsilon_s} +\frac{1}{\epsilon_s-\epsilon_t}  +
\frac{\epsilon_s}{\Delta}& 
  \frac{1}{\epsilon_s}
-\frac{1}{\epsilon_s-\epsilon_t}-\frac{\epsilon_t}{\Delta}
\end{array}
\right),
\end{eqnarray}
which can be diagonalized by the Harrison-Perkins-Scott matrix
\cite{Harrison:2002er}  
\begin{eqnarray}
U_{HPS}=\left(\begin{array}{ccc}\sqrt{2/3}&1/\sqrt{3}&0\\
-1/\sqrt{6}&1/\sqrt{3}&-1/\sqrt{2}\\
-1/\sqrt{6}&1/\sqrt{3}&1/\sqrt{2}\end{array}\right).
\end{eqnarray}
Recall that the charged lepton sector is already diagonal in this
basis and therefore, $U_{HPS}$ gives the PMNS mixing matrix with the
predicted TBM form. 
The resulting neutrino mass spectrum reads   
\begin{eqnarray}
U_{HPS}^T~ \tilde{\mathcal{M}}^{\nu}~
U_{HPS}=-\frac{\tilde{m}}{\iota}
\left(\begin{array}{ccc}
\frac{1}{\epsilon_s+\epsilon_t}&0&0\\
0&\frac{1}{\epsilon_s}&0\\
0&0&\frac{1}{\epsilon_t-\epsilon_s}\end{array}\right),
\end{eqnarray}
implying the neutrino mass-squared differences 
\begin{eqnarray}
\label{dosuno}
\Delta m_{21}^2&\equiv &
|m_2|^2-|m_1|^2=\left|\frac{\tilde{m}}{\iota \epsilon_s}\right|^2
\left[1-\frac{1}{(1+r)^2}\right],
\\
\label{tresuno}
\Delta m_{31}^2&\equiv &
|m_3|^2-|m_1|^2=\left|\frac{\tilde{m}}{\iota \epsilon_s}\right|^2
\left[\frac{4r}{(1-r^2)^2}\right],
\end{eqnarray}
where $r\equiv \epsilon_t/\epsilon_s$. 
From Eq. (\ref{dosuno}) we see
that $\Delta m_{21}^2$ is positive, as conventionally assumed, 
for $r<-2$ or $r>0$. 
(For $-2<r<0$ we would have to exchange the ordering  of the first two
neutrinos, thus ruining the TBM prediction.) 
Hence, the neutrino spectrum is normal
($\Delta m_{31}^2>0$) 
for $r>0$ and inverted ($\Delta m_{31}^2<0$) 
for $r<-2$ (see Eq. (\ref{tresuno})).
There are three solutions to 
Eqs. (\ref{dosuno}) and (\ref{tresuno}) 
reproducing the observed mass-squared differences, 
$\Delta m_{21}^2 \approx 7.67\times 10^{-5}~ \mathrm{eV}^2$ 
and $\Delta m_{31}^2 \approx 2.46~(-2.37) \times
10^{-3}~\mathrm{eV}^2$ for normal (inverted) 
hierarchy~\cite{GonzalezGarcia:2007ib}, 
in the allowed $r$ range, 
\begin{equation}
r\approx -2.01,0.79,1.20~.
\end{equation}
The other solution $r\approx -1.99$ does not give the correct mixing pattern 
and is therefore ignored.  
However, both, the normal ($r=0.79,1.20$) 
and the inverted ($r=-2.01$) mass hierarchy, can be realized 
in these models, with similar phenomenology in either case. 
On the other hand, 
the correct scale of neutrino masses is easily obtained 
varying the localization parameter $c_2$, which lies in the interval 
$-0.4 \lesssim c_2 \lesssim -0.2$ for $c_{1,3}$ values 
giving the $\tau$ mass and $|\epsilon_{t,s}| \sim
\mathcal{O}(10^{-2}-10^{-1})$.

These results receive three types of corrections. First, there are
bulk lepton KK modes with masses $\sim$ TeV which mix with the
zero modes. This mixing is small for leptons localized 
near the UV brane, 
and therefore the modifications they induce on the fermion masses 
and mixings are small too. However since the inter-generational mixing 
is large in the lepton sector, it is important to check that no large 
LFV is introduced. The second source of corrections is
related to the perturbative treatment of the Higgs effects. 
This is justified for the scales allowed by EWPT, 
but in GHU models we can actually test how good this 
approximation is because in this case it is possible to get a solution  
to all orders in the Higgs v.e.v.. 
These two types of corrections, which do not significantly modify the 
picture drawn above, 
are studied in the next two subsections. 
Finally, we have only included the LO 
$A_4 \times Z_8$ breaking terms. Higher orders, although suppressed 
by extra powers of $1/\Lambda^{(\prime)}$, could destabilize the 
TBM mixing pattern and introduce new sources of LFV. 
We will consider these higher order corrections in the following 
section.

\subsection{Inclusion of massive KK modes} 

The lepton mass Lagrangian contains a Dirac part that includes the
Yukawa Lagrangian plus the KK mass terms,
\begin{equation}
\mathcal{L}_D =
\mathcal{L}_Y-\sum_{n\ge 1}\left[m_n^{l}
\overline{l}_L^{(n)}l_R^{(n)}
+m_n^{\tilde{l}_{1/2}}\overline{\tilde{l}}_{\frac{1}{2}L}^{(n)}
\tilde{l}_{\frac{1}{2}R}^{(n)}+
m_n^{\tilde{l}_{-3/2}}
\overline{\tilde{l}}_{-\frac{3}{2}L}^{(n)}
\tilde{l}_{-\frac{3}{2}R}^{(n)}\right. 
+ 
\left.m_n^{\nu^{\prime}}\overline{\nu}^{\prime (n)}_L
\nu^{\prime (n)}_R+m_n^{e^{\prime}}
\overline{e}_L^{\prime (n)}e^{\prime (n)}_R+\mathrm{h.c.} \right],
\end{equation}
where the $SU(2)_L$ doublets with hypercharges $\frac{1}{2}$ and
$-\frac{3}{2}$ are denoted by $\tilde{l}_{\frac{1}{2}}$ and 
$\tilde{l}_{-\frac{3}{2}}$, respectively, and the 
SM-like (hypercharge $-\frac{1}{2}$) doublets 
which participate from all $SO(5)$ multiplets by $l$. 
Obviously, $\mathcal{L}_Y$ also 
includes Yukawa couplings with the massive KK modes. 
The Dirac mass Lagrangian can be written in matrix form 
\begin{equation}
-\mathcal{L}_D=
\overline{e}_L \mathcal{M}_D^ee_R
+\overline{\nu}_L\mathcal{M}^{\nu}_D\nu_R
+\sum_{n\ge 1}m_n^{\tilde{l}_{1/2}}\overline{\widetilde{X}}_{L}^{(n)}
\widetilde{X}_{R}^{(n)}+\sum_{n\ge 1}m_n^{\tilde{l}_{-3/2}}
\overline{\widetilde{Y}}_{L}^{(n)}\widetilde{Y}_R^{(n)}+\mathrm{h.c.},
\end{equation}
where we have grouped together the charge $-1$ leptons into $e_{L,R}$ and
the neutral ones into $\nu_{L,R}$.~\footnote{In this subsection we use
the same calligraphic notation to denote matrices although they
have a larger size here because they also include massive KK modes.}
The UV brane term, Eq. (\ref{UVMajorana:mass}), induces a
Majorana mass term that now involves all KK modes of the RH neutrinos 
\begin{eqnarray}
-\mathcal{L}_M=
\frac{1}{2}\overline{\nu}_R^c
\mathcal{M}^{\nu}_M \nu_R+\mathrm{h.c.}~.
\end{eqnarray}
The mass Lagrangian is diagonal for the charge $+1$ and $-2$ sectors
but not for the charge $-1$ and neutral ones. 
However, it is still true that it is family diagonal
except for the terms involving the Majorana neutrino masses. 
Thus, although we have now to diagonalize the charged lepton mass term, 
this diagonalization does not mix different generations and 
then does not introduce flavor changing neutral currents (FCNC). The
corresponding modification of the diagonal $Z$ couplings 
is proportional to the charged lepton masses and therefore
relatively small~\cite{delAguila:2000kb}.
On the other hand, the neutrino mass matrix 
\begin{eqnarray}
\mathcal{M}^{\nu}=\left(\begin{array}{cc}0&\mathcal{M}_D^{\nu}\\
\left(\mathcal{M}_{D}^{\nu}\right)^T&\mathcal{M}_M^{\nu}\end{array}\right)
\end{eqnarray} 
is not family diagonal, 
and
the required rotation could in principle induce modifications of the TBM 
mixing and introduce dangerous non-diagonal 
couplings between the SM charged leptons and the neutrino KK modes 
of mass $\sim$ TeV, implying large LFV processes 
at the loop level. 
We have numerically checked that neither of these two possibilities is
actually realized. The inclusion of massive KK modes does not
appreciably modify the TBM mixing pattern and furthermore, although
there are non-negligible charged couplings between the SM charged
leptons and the neutrino KK modes, they are, to an excellent
approximation, family diagonal, \textit{i.e.} if the coupling $e
N$ is sizable for some heavy $N$, then the couplings $\mu N$ and
$\tau N$ are extremely suppressed. This can be easily
understood observing that flavor violation (and also 
light neutrino masses and thus TBM mixing) is induced by 
the corresponding Majorana mass, which being localized at the UV
brane is much larger than the TeV scale. (For a detailed discussion
of the effect on neutrino masses and mixing see~\cite{Huber:2003sf}.) 

\subsection{Exact Higgs treatment}

GHU models like the one we are considering are among the
best motivated models with warped extra dimensions, due to the extra
protection of the Higgs potential. They are also interesting because
they allow us to solve the bulk equations of motion in the presence of
a bulk Higgs. We can perform a
field redefinition identical to a gauge transformation which locally
removes the Higgs from the action, except at one of the branes. 
Then, the Higgs does not enter in the bulk action for rotated fields 
but only as a boundary condition, which can be implemented
numerically. 
We can, therefore, 
compute non-linear effects of the Higgs due to its Goldstone
boson nature. These effects are typically small for the values of the KK
scale allowed by EWPT, but this exact treatment will allow us to 
test our approximation. 
Besides, we can also include the UV localized Majorana masses as exact 
boundary conditions, instead of perturbatively. 

The field transformation that removes the
Higgs locally except at the IR brane is identical to a gauge
transformation with gauge parameter 
\begin{equation}
\rho(z,v_H) = \exp \left[ i 
\frac{\sqrt{2}g_5 v_H T_C^4}{\sqrt{R(R^{\prime\,2}-R^2)}} \int_R^z dz^\prime
z^\prime \right]
\approx 
\exp \left[ i g v_H T^4_C \sqrt{\log(R^\prime/R)/2} \left(
\frac{z^2-R^2}{R^\prime}\right)\right].
\end{equation}
This is not an actual gauge transformation because this gauge
parameter does not satisfy the corresponding boundary conditions, 
but it eliminates the Higgs boson locally except at the IR brane.
The bulk action for the rotated fields
\begin{equation}
\zeta^\prime = \rho \zeta 
\end{equation}
is then free of the Higgs v.e.v. $v_H$, 
and it can be solved analytically as we did
before. The boundary conditions at the UV brane remain the same, since
$\rho(R)=1$. However, 
the boundary conditions at the IR brane in Eq.~(\ref{ir}) 
apply to the original fields and when written in
terms of the rotated ones, they will explicitly include the Higgs
effects. Note that the physical modes will now participate from all 
multiplets, not only from those mixed by localized terms but from 
those mixed by the Higgs, too. 
This makes the corresponding boundary conditions much more
challenging. Also note that, since we are imposing now as boundary
conditions the UV localized Majorana masses, we necessarily have to deal
with all three generations simultaneously in the neutrino expansion. 
For instance, once we impose the UV boundary conditions, 
the Higgs dependent IR boundary conditions give a system of 8
equations with 8 unknowns (per family) for the charge $-1$ leptons and two
independent systems of 24 equations with 24 unknowns for the neutral
ones (due to the Majorana boundary condition the three families mix
and the corresponding system of 24 equations with 24 complex unknowns
splits into real and imaginary parts, as discussed in
Appendix~\ref{appendix:majorana}).
Requiring a non-trivial solution of the corresponding systems fixes
the mass of the physical states and determines all unknowns in terms
of one normalization constant, which is then fixed by the normalization
condition involving all relevant multiplets.
The exact expression for these boundary conditions are too large to
be included here but we have checked that the
masses of the charged and neutral leptons (for simplicity we have
neglected inter-generational mixing) are in 
excellent agreement with those obtained with a
perturbative treatment of Higgs and UV Majorana mass effects. 

\section{Higher order effects\label{higher:orders}}

We have seen in the previous section that a global $A_4$ symmetry
can naturally explain the observed lepton masses and TBM neutrino 
mixing at LO in the breaking of this discrete symmetry along the 
appropriate direction. 
The zero mode pattern remains almost unchanged 
when lepton KK modes or bulk Higgs effects are included. 
Furthermore, this global symmetry provides an extra level of
flavor protection that makes the model compatible with experimental
data despite the large number of new particles. 
The nearly exact realization of TBM mixing, the 
very precise cancellation of flavor violations and the $\tau$ mass 
preference for a not too small value of $v^\prime/\Lambda^\prime$ (or
alternatively a large degree of compositeness) 
must be also verified at higher orders in the global symmetry breaking. 
The structure of higher order contributions is greatly simplified 
because $\phi$ ($\phi^\prime$) preserves a $Z_2$ ($Z_3$) subgroup of 
$A_4$~\cite{Csaki:2008zd}. 
In practice, this means that 
\begin{equation}
\langle \phi \rangle ^3 \sim \langle \phi \rangle, \quad
\langle \phi^\prime \rangle ^2 \sim 1+\langle \phi^\prime \rangle,
\end{equation}
where $\sim$ means that they have the same $A_4$ transformation 
properties. Hence, only operators with one or two powers of $\phi$ 
on the UV brane and operators with none or one power of $\phi^\prime$ 
on the IR brane give rise to independent flavor structures. 
The allowed operators are further constrained by the $Z_8$ symmetry. 

The Majorana neutrino masses on the UV brane already have 
terms with none and one power of $\phi$, so the only new
structure comes from operators with two powers of $\phi$. 
The lowest order contribution compatible with $Z_8$ has of the form 
\begin{equation}
\frac{\eta\phi^2}{\Lambda^3} 
\overline{\nu}^{\prime\,c}_{2R} 
\nu^\prime_{2R}
+ \mathrm{h.c.} \to 
\overline{\nu}^{\prime\,c}_{2R} 
\begin{pmatrix}
\delta_1+\delta_2+\delta_3 & 0 & 0 \\
0 &\delta_1+\omega \delta_2 +\omega^2 \delta_3 & 0 \\
0 & 0 & \delta_1+\omega^2 \delta_2+\omega \delta_3
\end{pmatrix} 
\nu^\prime_{2R}
+ \mathrm{h.c.},
\end{equation}
with $\delta_i\sim v_\eta v^2/\Lambda^3$ arbitrary. 
The boundary coupling between the $\zeta_1$ and $\zeta_3$ bi-doublets 
gets new structures from terms with one or two powers of $\phi$. 
The latter gives a similar contribution to the previous one for neurinos, 
whereas the former gives a $2-3$ mixing, 
\begin{equation}
\left[\frac{\eta \phi}{\Lambda^2}  +
\frac{\phi^2}{\Lambda^2}\right] 
\overline{l}_{1L} 
l_{3R}
+
\mathrm{h.c.} 
\to
\overline{l}_{1L} 
\begin{pmatrix} 
 \rho_1 +\rho_2+\rho_3& 0 & 0 \\
0 &\rho_1+\omega\rho_2+\omega^2\rho_3 & \gamma_1 \\
0 & \gamma_2 & \rho_1+\omega^2\rho_2+\omega \rho_3
\end{pmatrix}
l_{3R}
+\mathrm{h.c.}, 
\end{equation}
where $\rho_i \sim v^2/\Lambda^2$ and $\gamma_i \sim v_\eta v/\Lambda^2$.

Let us discuss now the terms on the IR brane. 
The leading term mixing $\zeta_1$ with $\zeta_2$
contains no power of $\phi^\prime$, so the only new structure
corresponds to one factor of $\phi^\prime$. 
The first such term comes at order $1/\Lambda^{\prime\,3}$, 
due to the $Z_8$ symmetry. 
At this order we have
\begin{equation}
\left[\frac{\eta^{\prime\,\ast\,2} 
\phi^{\prime}}{\Lambda^{\prime\, 3}} 
+\frac{\eta^{\prime} 
|\phi^\prime|^2}{\Lambda^{\prime\, 3}} 
+
\frac{\eta^{\prime\,\ast} 
\phi^{\prime\,\dagger\,2}}{\Lambda^{\prime\, 3}} 
\right]
\left(
\overline{l}_{1L} 
l_{2R}
+\overline{\tilde{l}}_{1L} \tilde{l}_{2R}\right)
+ \mathrm{h.c.} 
\to 
\overline{l}_{1L} 
\begin{pmatrix}
\epsilon_1 & \epsilon_2 &\epsilon_3 \\
\epsilon_3 & \epsilon_1 &\epsilon_2 \\
\epsilon_2 & \epsilon_3 &\epsilon_1 
\end{pmatrix}
l_{2R}
+
[l_{1,2}\to \tilde{l}_{1,2}]+\mathrm{h.c.},
\label{12mixing:nlo:before:rotation}
\end{equation}
where 
$\epsilon_1\sim v_\eta^\prime v^{\prime\,2}/\Lambda^{\prime\,3}$ and
$\epsilon_{2,3}$ get contributions $\sim v_\eta^\prime
v^{\prime\,2}/\Lambda^{\prime\,3}$ and $\sim
v_\eta^{\prime\,2} v^{\prime}/\Lambda^{\prime\,3}$, and similarly 
for $\overline{\nu}^\prime_{1L} \nu^\prime_{2R}$. 
Finally, the coupling between $\zeta_3$ and $\zeta_\alpha$ is not
modified by higher order terms, because we already have a term with one
power of $\phi^\prime$ and the singlet contribution cannot result from 
a singlet structure under $A_4$. 
No further structures are generated at higher orders. 

Therefore the higher order effects in the $A_4$ breaking can 
be summarized, after the rotation $\zeta_k \to \Omega\ \zeta_k$ in 
Eq. (\ref{rotation}), by the following replacements 
\begin{equation}
\hat{\theta}_M \to \hat{\theta}_M +
\begin{pmatrix}
\delta_1 & \delta_3 & \delta_2 \\
\delta_3 & \delta_2 & \delta_1 \\
\delta_2 & \delta_1 & \delta_3
\end{pmatrix} \label{newtheta}
\end{equation}
for the Majorana masses, 
\begin{equation}
x_l \to x_l + X_l,
\end{equation}
with 
\begin{equation}
X_l=\Omega^\dagger \begin{pmatrix} 
 \rho_1 +\rho_2+\rho_3& 0 & 0 \\
0 &\rho_1+\omega\rho_2+\omega^2\rho_3 & \gamma_1 \\
0 & \gamma_2 & \rho_1+\omega^2\rho_2+\omega \rho_3
\end{pmatrix}
\Omega \label{newxl}
\end{equation}
for the mixing between the SM LH doublets in $\zeta_1$ and $\zeta_3$, and
\begin{equation}
y_{b,s} \frac{v^\prime}{\Lambda^\prime} 
\to y_{b,s} \frac{v^\prime}{\Lambda^\prime}  + 
\begin{pmatrix} 
\epsilon_1^{b,s}+\epsilon_2^{b,s}+\epsilon_3^{b,s} & 0 & 0 \\
0 & \epsilon_1^{b,s}+\omega \epsilon_2^{b,s}+ \omega^2\epsilon_3^{b,s} \\
0 & 0 & \epsilon_1^{b,s}+\omega^2\epsilon_2^{b,s}+
\omega \epsilon_3^{b,s} 
\end{pmatrix}
\label{12mixing:nlo:after:rotation}
\end{equation} 
for the mixing between the bi-doublets or the singlets 
in $\zeta_1$ and $\zeta_2$. 
The IR terms remain diagonal whereas the UV terms receive non-diagonal
corrections. All three effects are a source of violation of TBM and
the non-diagonal $X_l$ a source of FCNC for the charged leptons. 
This implies some constraint on
$v^{(\prime)}_{(\eta)}/\Lambda^{(\prime)}$ 
that will be discussed in the next section.

\subsection{A comment on brane kinetic terms}

We have neglected so far in our discussion the effect of brane
kinetic terms (BKT). These are generated by quantum corrections
and therefore cannot be set to zero at arbitrary
scales~\cite{Georgi:2000ks}. The
global symmetries of our model, however, strongly constrain them. 
In particular, all possible BKT are proportional to the identity 
at leading order in $A_4$ and $Z_8$ breaking, 
except those involving $\zeta_\alpha$ fields, 
which are diagonal but flavor dependent. 
Corrections to this pattern are of order
$v^2/\Lambda^2$, where $v$ and $\Lambda$ stand here for any 
$v,v_\eta,v^\prime, v^\prime_\eta$ and $\Lambda,\Lambda^\prime$,
respectively. 
Since at leading order there is no flavor violation in
the charged lepton sector, these flavor dependent (but diagonal) BKT 
do not generate FCNC. Once higher order terms are included, there is a
small flavor violation in the charged lepton sector and
therefore the flavor dependent BKT will induce FCNC. 
Higher order contributions to BKT breaking $A_4$ 
are also a potential source of flavor violation. However, the
$A_4$ and $Z_8$ symmetries ensure that the
$v^{\prime\,2}/\Lambda^{\prime\,2}$ corrections are diagonal. 
Therefore, their effect is subleading and we will disregard 
them here. 
Hence, we include higher order
effects in the localized mass terms but not in the BKT.
The effects of diagonal BKT are well-known (see for
instance~\cite{BKT}). 
They
do not change the functional dependence of the fermion zero modes and
only affect their normalization. The ones leading to flavor violation
are the BKTs for the RH component of the $SO(4)$
singlet in $\zeta_\alpha$ (all the other ones are proportional to the
identity, up to tiny corrections). The corresponding BKT can be
written as 
\begin{equation}
\delta S=\int d^4x dz a^4 \delta(z-R^\prime)
R^\prime \kappa^\alpha \overline{e}^\prime_{\alpha R} i \cancel{D}
e^\prime_{\alpha R} + \ldots, \label{S:BKT}
\end{equation}
where $\kappa^\alpha $ is a dimensionless coefficients parametrizing
the BKTs. The fermion zero modes for the RH charge $-1$ leptons have the
same functional form as in the absence of BKT, Eqs. 
(\ref{eR3:zeromode}) and (\ref{eRa:zeromode}), except for the
normalization that is now
\begin{equation}
\rho_\alpha^\mathrm{BKT}= \rho_\alpha+ \kappa^\alpha f^2_{-c_\alpha}.
\end{equation}
Note that $f^2_{-c} \ll 1$ for $c\lesssim -0.4$ so this effect in the
normalization can be only relevant for the tau lepton. A more significant
effect regarding flavor is that the covariant derivative in
Eq. (\ref{S:BKT}) contains the KK expansion of the corresponding
gauge bosons. This implies the following flavor dependent coupling of
the fermion zero modes to the gauge boson KK modes
\begin{equation}
\delta S=\int d^4x \Big[g \sqrt{R \log (R^\prime/R)}
\kappa^\alpha \frac{f^2_{-c_\alpha}}{\rho_\alpha^\mathrm{BKT}}
f_n^A(R^\prime) \Big] \bar{e}_{\alpha R} \cancel{A}_n e_{\alpha R} + \ldots, 
\label{S:BKT:couplings}
\end{equation}
which has to be added to the bulk contribution.
We have used again the tree level matching of the
coupling constant Eq. (\ref{matching}) and assumed a KK expansion
of the gauge bosons
\begin{equation}
A_\mu(x,z)= \sum_n f^A_n(z) A_\mu^n(x),
\end{equation}
with $A_\mu(x,z)$ a generic gauge field (we have left the group
structure implicit). After the inclusion of higher
order terms in the brane mass terms discussed in the previous section,
the charged lepton sector is no longer flavor diagonal in the current
eigenstate basis. The rotation of the RH charged leptons required to
go to the physical basis will then induce flavor violating couplings
to the gauge boson KK modes. Recall however that the charged lepton
mass hierarchy is obtained by means of the localization of the RH
charged lepton zero modes. This implies that the RH rotation to go to
the physical mass is strongly hierarchical and therefore the FCNC
induced by the BKT suppressed by the charged lepton masses. 
We have indeed numerically checked that BKTs do not impose any
significant constraint in the model and we will therefore neglect them
in the discussion about electroweak and flavor constraints in the next
section. 

\section{Electroweak and flavor constraints\label{constraints}} 

Once we have classified all possible higher order terms
in the $A_4\times Z_8$ breaking expansion, 
we can discuss their effects on the leptonic mixing, 
\textit{i.e.} departures from TBM mixing, as well as
LFV. All three new flavor structures,
Eqs. (\ref{newtheta}-\ref{12mixing:nlo:after:rotation}), 
are a source of departure from TBM mixing; whereas LFV is 
mostly affected by Eq. (\ref{newxl}). Given the large number of
parameters in our model, it is difficult to establish detailed bounds
on each one. However, there are some general tendencies that are easy
to understand. We have performed a detailed scanning to test these
tendencies. The main conclusion is that a large region of parameter space
is allowed by all current electroweak and flavor data for an IR scale
$1/R^\prime= 1.5$ TeV, provided $v/\Lambda$ is not too large
($\lesssim 0.1$) and the LH charged leptons are close to the UV brane
($c_{3,1} \gtrsim 0.5$). This conclusion might seem a bit surprising,
given previous studies of LFV in models with warped extra
dimensions~\cite{Agashe:2006iy}. The reason our model works so well
regarding LFV is a combination of two types of flavor protection. The
first one is the protection provided by the $A_4$ symmetry, which in
simpler models with warped extra dimensions is enough to ensure
agreement with experimental data~\cite{Csaki:2008qq}. In our case,
due to the richer structure imposed by GHU models, this protection is
not enough. This is where the second layer of flavor protection
enters. Our model naturally falls in the optimal configuration
discussed in~\cite{Agashe:2009tu}. The custodial symmetry, together
with a LR symmetry originally proposed to protect the 
$Z \bar{b}_L b_L$ coupling~\cite{Agashe:2006at}, and the splitting of
the SM fields in two separate sectors ($\zeta_{1,2}$ and
$\zeta_{3,\alpha}$ in our case) reduce LFV in our model to values 
compatible with current data, despite the low scale of new
physics. 
\footnote{Recent analysis of LFV in 4D
supersymmetric models with an $A_4$ symmetry can be found
in~\cite{LFV:A4}.} 

We must require for the model to be realistic 
that it satisfies all experimental constraints. 
We can classify them in four types: those from EWPT,  
which we will estimate requiring small deviations 
from the SM tree-level couplings; 
limits on LFV processes which can proceed at tree-level; 
bounds on LFV processes which are banished to higher 
orders; 
and contraints from neutrino oscillations. 
The first three types of restrictions are mainly 
related to the heavy spectrum, 
whereas the latter one depends more directly on 
the discrete flavor symmetry breaking. 
Thus, although it involves less precisely determined 
parameters, it does restrict the model. 
The following phenomenological analysis 
must be understood as an existence 
proof. 
A refined analysis, which is outside the scope of this paper, 
should consistently include all contributions to a 
given order. 
We have done this for tree-level processes, 
but not for one-loop contributions which have been only 
estimated with the typically larger amplitude baring, 
for instance, possible cancellations. 
On the other hand, we have not considered one-loop 
corrections to $Z \bar{e} \mu$~\cite{Tommasini:1995ii}. 
A detailed study of this type of constraints 
will be presented elsewhere, for they require a precise 
enough (numerical) treatment of fermion mixing to 
recover the proper behaviour of the different contributions, 
and then of decoupling \cite{Appelquist:1974tg}.   
The restrictions we explicitly consider are:
\begin{itemize} 
\item \textbf{Electroweak precision tests}. 
We have required the gauge couplings of the SM charged leptons to be 
in agreement with the SM prediction within 2 per mille
accuracy~\cite{Csaki:2008qq}, both for neutral 
$Z \bar{l}_\alpha l_\alpha $ and charged $W \bar{l}_\alpha\nu_\alpha$ 
currents. 
This is typically the present limit on the mixing of the 
electroweak gauge bosons with new resonances~\cite{Amsler:2008zzb}, 
and on the square of the SM lepton mixing with heavier 
vector-like fermions~\cite{delAguila:2008pw}.  
\item \textbf{Tree-level LFV}. We have included the most 
relevant constraints following~\cite{Agashe:2006iy}.
Explicitly,  we
have studied the decays $\mu\to e^-e^+e^-$, $\tau\to\mu^-\mu^+\mu^-$,
$\tau\to e^-e^+e^-$, $\tau\to \mu^-e^+e^-$, $\tau\to\mu^- e^+ \mu^-$
and the $\mu-e$ nuclear conversion rate. The tri-lepton decays $l\to
l_1 \bar{l}_2l_3$ are mediated by LFV tree-level couplings to the 
physical $Z$ gauge boson and its KK excitations. 
(The effects due to fermion
mixing are negligible).~\footnote{Higgs mediated
  contributions~\cite{Agashe:2009di} are suppressed by the $A_4$
  symmetry and the SM lepton masses, and then very small in this class
  of models.}  
At low energies,
these contributions can be parameterized by the following effective Lagrangian, 
\begin{eqnarray}
-\mathcal{L}_{\mathrm{eff}}&=&
\frac{4G_F}{\sqrt{2}}\left[g_3^{\alpha\beta}\left(\bar{l}^{\beta}_R 
  \gamma^{\mu}
  l^{\alpha}_R\right)
\left(\bar{l}_R^{\beta}\gamma^{\mu}l_R^{\beta}\right)
+g_4^{\alpha\beta}\left(\bar{l}^{\beta}_L
  \gamma^{\mu}
  l^{\alpha}_L\right)
\left(\bar{l}_L^{\beta}\gamma^{\mu}l_L^{\beta}\right)\right.\nonumber\\  
&&+\left.g_5^{\alpha\beta}\left(\bar{l}^{\beta}_R \gamma^{\mu}
l^{\alpha}_R\right)\left(\bar{l}_L^{\beta}\gamma^{\mu}l_L^{\beta}\right)
+g_6^{\alpha\beta}\left(\bar{l}^{\beta}_L
\gamma^{\mu}
l^{\alpha}_L\right)\left(\bar{l}_R^{\beta}\gamma^{\mu}l_R^{\beta}\right)
\right]+\mathrm{h.c.}, 
\end{eqnarray}
where $\alpha=e,\mu,\tau$. In terms of this effective Lagrangian, the
branching ratios for these decays read
\begin{eqnarray}
\mathcal{B}(\mu \to e e e)&=&2\left(|g_3^{\mu e}|^2+|g_4^{\mu
  e}|^2\right)+|g_5^{\mu e}|^2+|g_6^{\mu e}|^2,\nonumber\\ 
\mathcal{B}(\tau\to \mu\mu\mu)&=&\left\{2\left(|g_3^{\tau
  \mu}|^2+|g_4^{\tau \mu}|^2\right)+|g_5^{\tau \mu}|^2+|g_6^{\tau
  \mu}|^2\right\}\mathcal{B}\left(\tau\to e\nu\nu\right),\nonumber\\ 
\mathcal{B}(\tau\to e e e)&=&\left\{2\left(|g_3^{\tau e}|^2+|g_4^{\tau
  e}|^2\right)+|g_5^{\tau e}|^2+|g_6^{\tau
  e}|^2\right\}\mathcal{B}\left(\tau\to e\nu\nu\right),\nonumber\\ 
\mathcal{B}(\tau\to e e\mu)&=&\left\{|g_3^{\tau \mu}|^2+|g_4^{\tau
  \mu}|^2+|g_5^{\tau \mu}|^2+|g_6^{\tau
  \mu}|^2\right\}\mathcal{B}\left(\tau\to e\nu\nu\right),\nonumber\\ 
\mathcal{B}(\tau\to e\mu\mu)&=&\left\{|g_3^{\tau e}|^2+|g_4^{\tau
  e}|^2+|g_5^{\tau e}|^2+|g_6^{\tau
  e}|^2\right\}\mathcal{B}\left(\tau\to e\nu\nu\right). 
\end{eqnarray}
For the $\mu-e$ conversion rate we have applied the usual expression 
\begin{eqnarray}
B_{\mathrm{conv}}=\frac{2p_e E_e G_F^2m_{\mu}^3\alpha^3 |F_q|^2
  Z_{\mathrm{eff}}^4Q_N^2}{\pi^2 Z
  \Gamma_{\mathrm{capt}}}\left[|g_R^{\mu e}|^2+|g_L^{\mu e}|^2\right], 
\end{eqnarray}
where $g_{L,R}^{\mu e}$ are the corresponding off-diagonal
$Z\bar{e}\mu$ couplings, $G_F$ is the Fermi
constant and $\alpha$ the QED coupling strength, while the other terms
are atomic physics constants defined in \cite{Kuno:1999jp}. We 
shall use the current PDG~\cite{Amsler:2008zzb} 
bounds for the tri-lepton decays and the titanium bound
$B_{\mathrm{conv}}<6.1\times 10^{-13}$ from the SINDRUM II
experiment~\cite{Wintz:1998rp}  
for $\mu-e$ conversion.
\item \textbf{One-loop LFV}.
We have also considered the constraints on 
gauge boson~\cite{Csaki:2008qq} and Higgs~\cite{Agashe:2006iy} 
mediated amplitudes for $\mu \to e \gamma$. 
The charged boson contributions to this branching ratio read 
\begin{equation}
\label{GaugeBR}
\mathcal{B}_{G}(\mu\to
e\gamma)=\frac{3\alpha}{8\pi}
\left[\left|
\sum_V 
\sum_{i} 
U_{\mu i}^{VL\ast}U_{ei}^{VL} 
F_1\left(\frac{m_i^2}{M_V^2}\right)
\right|^2 + L \to R \right],
\end{equation}
where $V$ denotes the gauge boson running in the loop, 
including the $W_L$ zero mode and its lightest KK excitation, 
and the first $W_{R}$ KK mode (the charged gauge boson in $SU(2)_R$). 
The subscript $i$ indicates the massive fermion running in the loop, and
$U^{VL,R}_{e,\mu i}$ stand for the electron and muon couplings to
the corresponding gauge boson and heavy lepton (in units of $g/\sqrt{2}$). 
Finally, the function $F_1$ is given by 
\begin{eqnarray}
F_1(z)=\frac{1}{6\left(1-z\right)^4}\left(10-43z+78z^2-49z^3+4z^4
+18z^3\log z\right). 
\end{eqnarray}
There is a comparable contribution from neutral gauge boson 
exchange, typically of opposite
sign~\cite{delAguila:1982yu,Blanke:2007db}.  
The Higgs mediated branching ratio
reads~\cite{delAguila:1982yu,Agashe:2006iy} 
\begin{eqnarray}
\label{HiggsBR}
\mathcal{B}_H(\mu\to e\gamma)=\frac{3\alpha}{8\pi}
\left[\left|\sum_i \Lambda_{\bar{e}_L i_R} \Lambda_{\bar{i}_L\mu_R} 
\frac{v_H^2}{2m_\mu m_i} 
F_2\left(\frac{m_H^2}{m_{i}^{2}}\right)\right|^2+  
L \leftrightarrow R \right],
\end{eqnarray} 
where $\Lambda$ is the corresponding Yukawa matrix, $v_H\approx 246$ GeV, 
\begin{eqnarray}
F_2(x)=\frac{1}{\left(1-x\right)^3}\left(1-4x+3x^2-2x^2\log x\right),  
\end{eqnarray}
and the sum runs over the leptonic KK modes. 
The contributions in Eqs. (\ref{GaugeBR}) and (\ref{HiggsBR}) are of similar 
order when the mixing between light (SM) and heavy (vector-like) leptons, 
which is encoded in $U$ and $\Lambda$, respectively, 
is explicitly taken into account, 
despite the apparently large enhancement factor $v_H/m_\mu$ 
in the latter case~\cite{delAguila:1982fs}. 
We will use the current limit 
$\mathcal{B}(\mu\to e\gamma)<1.2\times 10^{-11}$~\cite{Amsler:2008zzb}, 
as well as the expected bound $\sim 10^{-13}$ from the on-going 
MEG experiment~\cite{Maki:2008zz} in the quantitative discussion below.
\item \textbf{PMNS matrix}.
We shall take the constraints on the 
PMNS mixing matrix from~\cite{GonzalezGarcia:2007ib}
\begin{equation}
|U|_{3\sigma}= \begin{pmatrix}
0.77 \to 0.86 & 0.50 \to 0.63 & 0.00 \to 0.22 \\
0.22 \to 0.56 & 0.44 \to 0.73 & 0.57 \to 0.80 \\
0.21 \to 0.55 & 0.40 \to 0.71 & 0.59 \to 0.82 \end{pmatrix}.
\end{equation}
\end{itemize} 

Let us discuss the scanning over the model parameters. 
Electroweak tests are generically satisfied for our choice of IR scale
$1/R^\prime=1.5$ TeV, as expected for UV localized light
fermions~\cite{Carena:2006bn} (with partial protection of 
universality for $Z$ couplings). The
constraints from tree-level LFV are also typically mild, due to the
double layer of flavor protection in our model. 
Among the processes considered only 
$\mu \to eee$ and $\mu - e$ conversion are close to current experimental
limits. 
In our general scan up to $\sim 70 \%$ and $\sim 51 \%$ of the
points pass the corresponding bounds, respectively. 
\footnote{$\mu-e$ conversion can be within the reach of 
projected experiments (see~\cite{LFV:A4}).}
The main constraint turns out to arise from $\mu \to e \gamma$. 
In general the new contributions are smaller for smaller values
of $v/\Lambda$ and relatively large values of $c_3$. 
On the other hand, the
departure from TBM mixing is somewhat sensitive to the value of 
$v^\prime/\Lambda^\prime$, decreasing with this ratio. 
Thus, in the following we fix $v^\prime/\Lambda^\prime=0.05$ 
to ensure a nearly correct neutrino mixing, passing 
$\sim 82\%$ of the points the PMNS test when varying 
the other parameters. 
These are randomly selected with
$0\leq v_\eta/\Lambda, v_\eta^\prime/\Lambda^\prime \leq
0.3$ and $c_3 \geq 0.5$. 
$v/\Lambda$ is computed from Eq. (\ref{dosuno}). 
\begin{figure}[!h]
{\includegraphics[width=0.475\textwidth,clip=true]{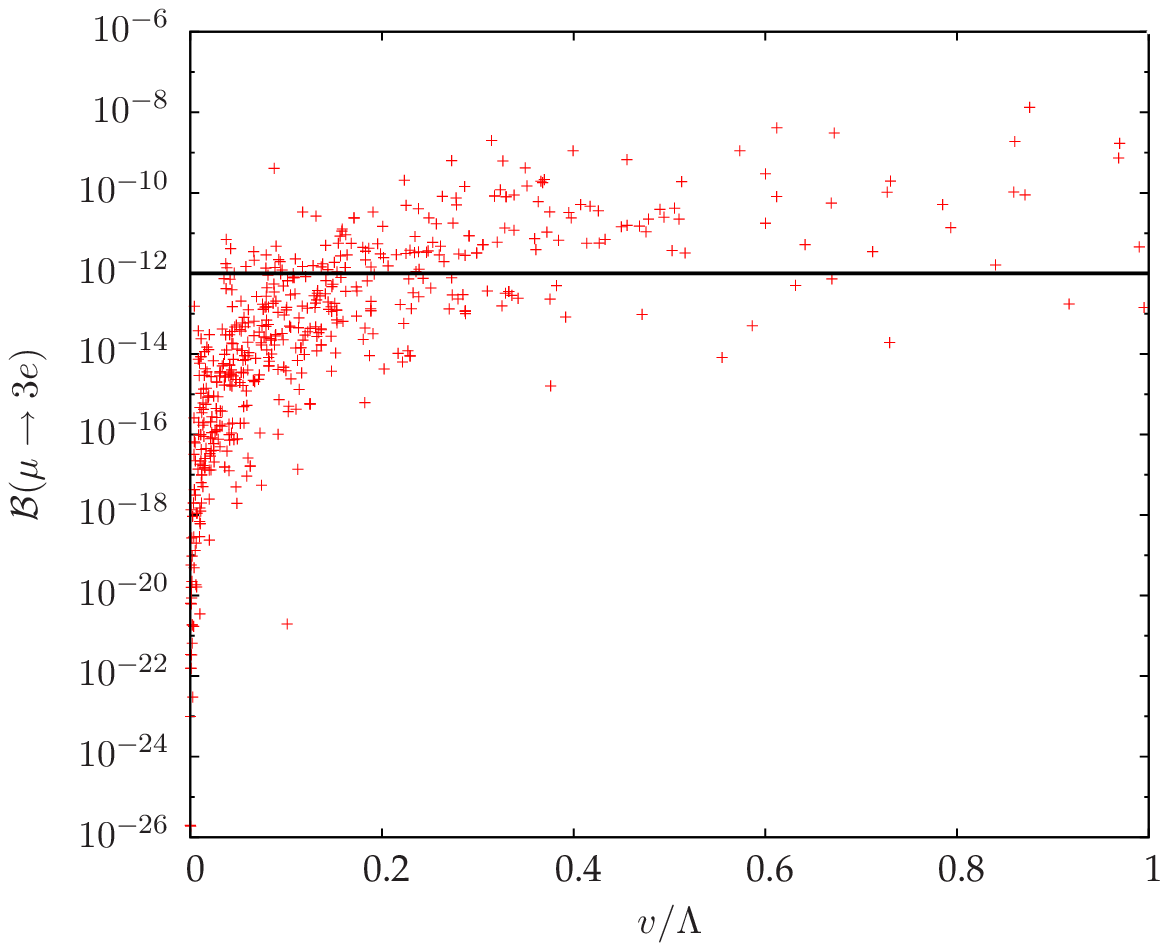}
\includegraphics[width=0.475\textwidth,clip=true]{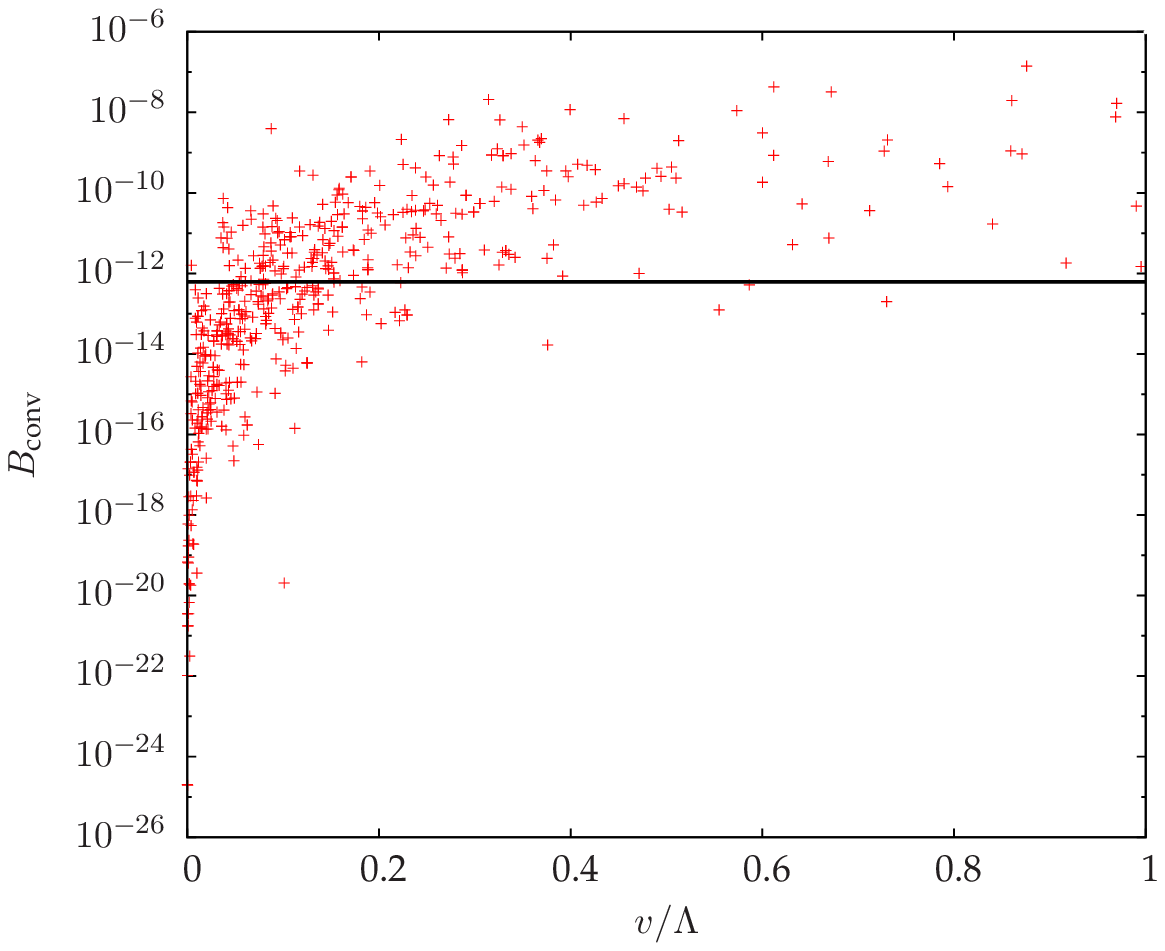}
\includegraphics[width=0.475\textwidth,clip=true]{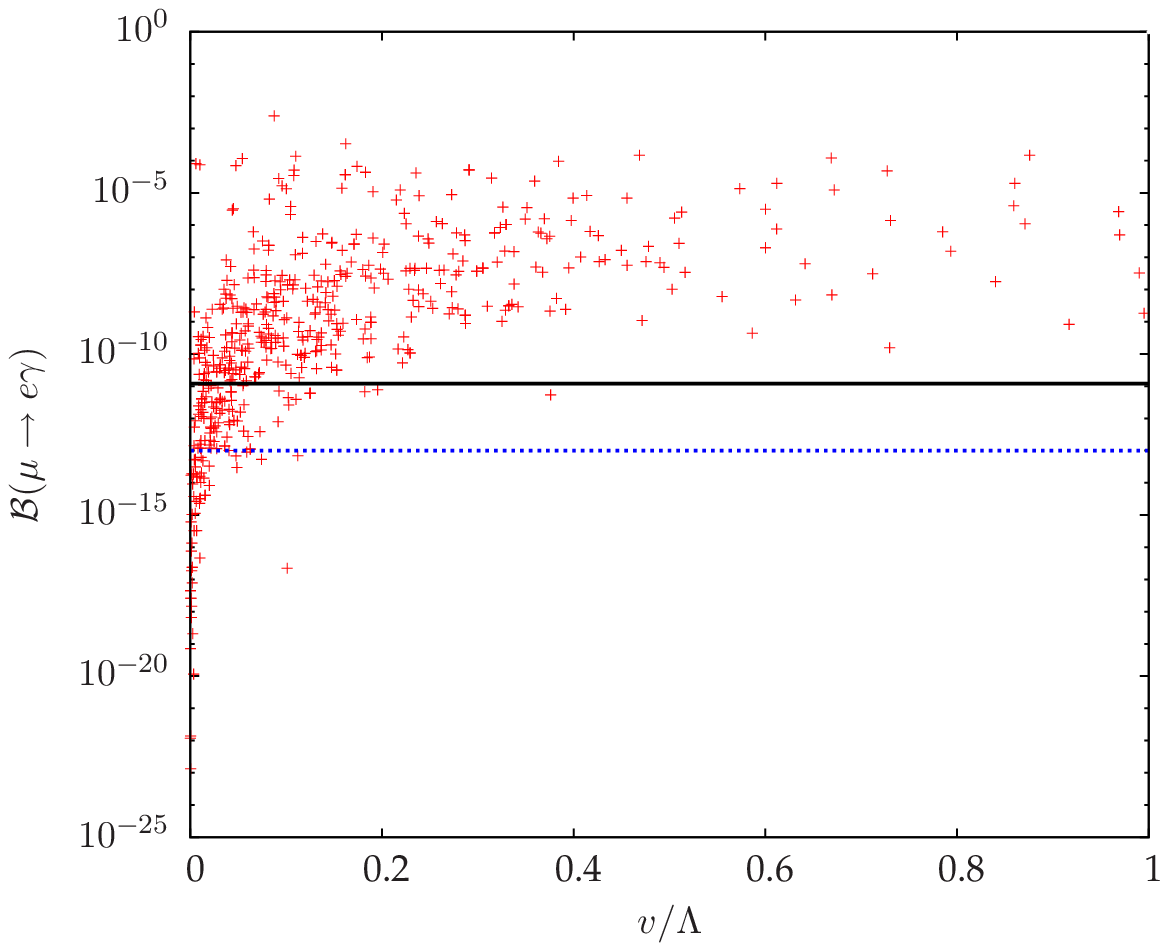}
\includegraphics[width=0.475\textwidth,clip=true]{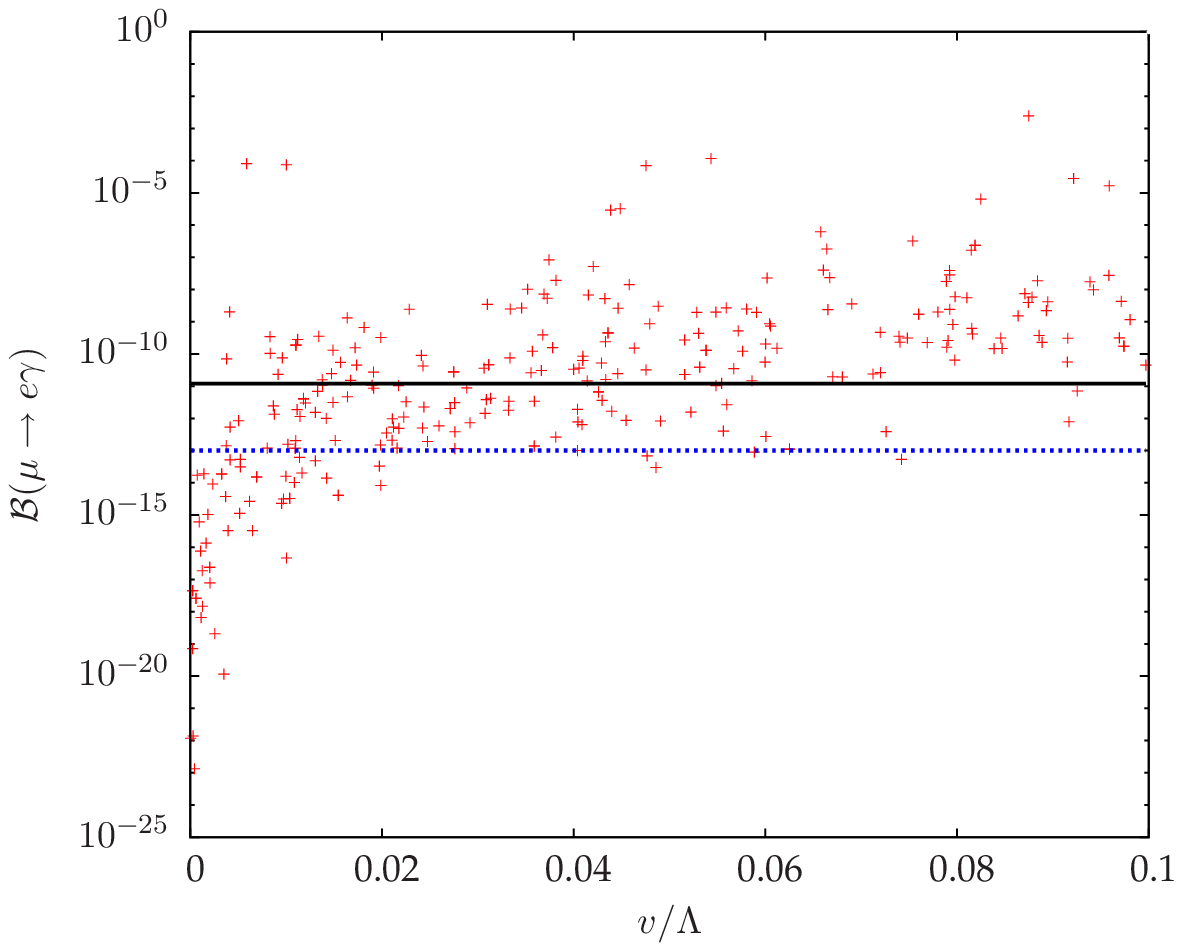}}
\caption{ 
LFV branching ratios as a function of $v/\Lambda$ for the scan 
described in the text. 
$\mu \to eee$ is plotted on the top-left panel, 
$\mu-e$ conversion in nuclei on the top-right one, and $\mu \to e
\gamma$ on the two lower panels 
(any $v/\Lambda$ value on the left panel and small $v/\Lambda$ values 
on the right one). 
The horizontal lines correspond to the current experimental upper 
bound (solid) and future sensitivity (dashed).}
\label{fig:LFV}
\end{figure}
In the Figure we show 
the most restricting observables as a function of
this ratio, together with the corresponding current 
experimental limit (solid line). 
These are $\mu \to eee$ (top-left panel), $\mu-e$ conversion
in nuclei (top-right panel) 
and $\mu \to e \gamma$ (lower panels, with the full range of 
$v/\Lambda$ on the left and only small $v/\Lambda$ values on the right), 
for which we have also drawn the expected sensitivity from MEG (dashed line). 
The extra flavor protection 
of tree-level mediated processes in the top panels relative to 
the one-loop $\mu \to e \gamma$ decay can be clearly observed in the Figure.
As we see, a large number of points
passes the different tests for relatively small 
$v/\Lambda$ values. If we restrict ourselves to $v/\Lambda \leq
0.05$, about $97\%$ of the points pass the $\mu \to eee$ and $\mu-e$
conversion tests, whereas $61\%$ satisfy the $\mu \to e \gamma$ bound 
(and only $28\%$ the expected MEG sensitivity). 
We collect in Table~\ref{scan:results} the 
percentage of points that satisfies all the experimental tests for 
different ranges of $c_3$ and $v/\Lambda$. 
Requiring $c_3\geq 0.55$ and $v/\Lambda \leq 0.05$ 
we find that $91\%$ of the points pass
all current experimental constraints ($53\%$ if we include the
projected MEG sensitivity on $\mu \to e \gamma$). 
For $v/\Lambda$ small enough, all tree-level LFV
effects are negligible, and the only (mild) constraint comes
from $\mu \to e \gamma$.
\begin{table}[!t]
\begin{center}
\begin{tabular}{|c|c|c|}
\hline
constraint & all tests &all tests + MEG 
\\
\hline
$c_3 \ge 0.5$, \quad $\frac{v}{\Lambda}\leq 0.05$ & $60\%$&$28\%$
\\
$c_3 \ge 0.55$, \quad $\frac{v}{\Lambda}\leq 0.15$ & $65\%$&$31\%$
\\
$c_3 \ge 0.55$, \quad $\frac{v}{\Lambda}\leq 0.05$ & $91\%$&$53\%$
\\\hline
\end{tabular}
\vspace{0.4cm}
\caption{Percentage of points that satifies all experimental tests 
(including the projected MEG sensitivity on the last column) 
for different parameter intervals.\label{scan:results}}
\end{center}
\end{table}
Note, however, that in our scans we have randomly selected 
order one values of the dimensionless couplings and fixed 
the global scale through the ratios
$v^{(\prime)}_{(\eta)}/\Lambda^{(\prime)}$. 
The unbalanced sensitivity of $\mu \to e \gamma$ forces 
the global scale to be small, and then all other effects are 
almost negligible, including deviations from TBM mixing. 
Of course, it is also possible that some couplings are 
accidentally larger than others, thus
inducing sizable corrections to some observables without
being excluded by the $\mu \to e \gamma$ limit. 
For example, if we set $v/\Lambda=0.5$ and all the coefficients 
of higher dimensional operators equal to zero except $\delta_2=8$ 
(well below its NDA estimate $\delta \lesssim 4\pi^2 x_\eta$ if 
$x_\eta \sim 1$), we obtain 
\begin{equation}
\sin \theta_{13}=0.18,
\end{equation}
with all other observables within experimental limits. Thus,
in our construction sizable departures from TBM mixing can 
be still compatible in with all other experimental constraints 
(although some fine-tuning might be necessary for large departures).

Our analysis shows that, in general, small values of 
$v/\Lambda$ and $v^{\prime}/\Lambda^{\prime}$ are preferred by 
lepton mixing and LFV observables.
We have already emphasized the correlation 
between $v^\prime/\Lambda^\prime$ 
and $c_\tau$ (the smaller the former, the larger $c_\tau$ has to be
in order to reproduce the $\tau$ mass, implying in turn 
a more composite $\tau_R$). This has important consequences regarding 
the spectrum in our model as
a larger $c_\tau$ value implies light modes.
The structure is very generic in this class of models. There is a
relatively light, almost degenerate bi-doublet
(two charge $-1$, one neutral and one charge $-2$ leptons) 
that mainly couples to $\tau_R$. 
This bi-doublet mostly lives in $\zeta_\tau$ (see Eq.~(\ref{multiplets})), 
which is light due to the assigned 
twisted boundary conditions~\cite{DelAguila:2001pu}. 
These four leptons can be very light and couple strongly to $\tau_R$ 
without being experimentally excluded because they are almost degenerate 
(see~\cite{Atre:2008iu} for a recent discussion of this phenomenon in
the quark sector). This degeneracy also dictates a very characteristic
collider phenomenology as we comment in the following.

\subsection{A numerical example and its collider implications}

As we have argued, it is very likely in this class of composite Higgs models 
that either 
there are LFV processes close to current experimental limits and 
then accessible at future flavor experiments or light leptonic resonances 
in an almost degenerate bi-doublet mainly coupled to $\tau_{R}$. 
Let us comment on an example predicting 
new leptonic resonances accessible at the LHC. 
The model is defined by the parameters 
\begin{eqnarray}
&& 
R^{-1}=10^{16}~\mathrm{TeV}, 
\quad
R^{\prime\,-1}=1.5 ~\mathrm{TeV}, 
\nonumber \\
&&
c_1= 0.65, \quad c_2= -0.19, \quad c_3= 0.57, 
\quad c_{e,\mu,\tau}= (-0.71,-0.54,0.49), 
\nonumber \\
&&
x_\eta=0.98, \quad x_\nu=1.28, \quad x_l=0.65,
 \\
&&
y_b=0.73, \quad y_s=0.44,
\quad y_b^\alpha =(1.10,-1.37,-0.36),
\quad y_s^\alpha =(-1.63,0.93,1.64), \nonumber \\ 
&&
\frac{v}{\Lambda}= 0.024, 
\quad
\frac{v_\eta}{\Lambda}= 0.04, 
\quad 
\frac{v^\prime}{\Lambda^\prime} = 0.05,
\quad 
\frac{v^\prime_\eta}{\Lambda^\prime} = 0.13, \nonumber
\end{eqnarray}
and random order one dimensionless coefficients for higher order
operators. 
The spectrum reproduces the observed pattern of charged lepton and 
neutrino masses and mixing angles. 
Above them, there is an almost degenerate
bi-doublet, whose matter content we denote by $N_{L,R}$,
$T^{1}_{L,R}$, $T^{2}_{L,R}$ and $Y_{L,R}$, all with masses $\approx
410$ GeV.  
Other fermionic resonances have masses 
$\gtrsim 3$ TeV.~\footnote{$N_{L,R}$ is 
a quasi-Dirac neutrino, for it has a tiny Majorana mass 
which is irrelevant for its collider phenomenology and 
will ignore in the following.} 
The bosonic resonances are all above $3.5$ TeV. 

The lepton couplings to the SM bosons can be written 
\begin{eqnarray}
\mathcal{L}^W&=& \frac{g}{\sqrt{2}} 
W^+_\mu 
\bar{\psi}^i_Q \gamma^\mu \Big[
V^{QL}_{ij} P_L +V^{QR}_{ij} P_R \Big]
\psi^j_{(Q-1)} 
+ \mathrm{h.c.}, 
\\
\mathcal{L}^Z&=& \frac{g}{2c_W}
 Z_\mu \bar{\psi}^i_Q \gamma^\mu \Big[
X^{QL}_{ij} P_L +X^{QR}_{ij} P_R - 2 s_W^2 Q \delta_{ij} \Big]
\psi^j_Q, 
\\
\mathcal{L}^H&=& -\frac{H}{\sqrt{2}} \bar{\psi}^i_Q Y^{Q}_{ij} P_R
\psi^j_Q + \mathrm{h.c.},
\end{eqnarray}
where $P_{L,R}$ stand for the chirality projectors
and $\psi_Q$ is the lepton of charge $Q$, when it exists. In our case
$Q=-2, -1, 0$ 
(and $+1$ if heavier modes are included). 

The relevant couplings, ignoring $e$ and $\mu$, read
\begin{equation}
|V^{(0)L}| \approx \begin{pmatrix}
0.71 & 0 & 0 \\
0 & 0.71 & 0.71 
\end{pmatrix},
\quad
|V^{(0)R}| \approx \begin{pmatrix}
0 & 0 & 0 \\
0.12 & 0.71 & 0.70
\end{pmatrix},
\end{equation}
\begin{equation}
|V^{(-1)L}|\approx \begin{pmatrix}
0 & 0.71 & 0.71
\end{pmatrix}^\mathrm{T},
\quad
|V^{(-1)R}|\approx \begin{pmatrix}
0.12 & 0.71 & 0.70
\end{pmatrix}^\mathrm{T},
\end{equation}
\begin{equation}
X^{(-1)L} \approx \begin{pmatrix}
-1 & 0 & 0 \\
0 & 0 & 1 \\
0 & 1 & 0
\end{pmatrix},
\quad
X^{(-1)R} \approx \begin{pmatrix}
0 & 0.16 & 0 \\
0.16 & 0 & 0.95 \\
0 & 0.95 & 0
\end{pmatrix},
\end{equation}
and 
\begin{equation}
Y^{(-1)}\approx \begin{pmatrix}
0 & 0 & 0 \\
0 & 0 & 0 \\
-0.39 & 0 & 0
\end{pmatrix},
\end{equation}
where a ``$0$'' entry means $\lesssim 10^{-2}$,  
and the order of charge $-1$ leptons is $\tau,T^1,T^2$.
These values are in good agreement with the expectations from degenerate
bi-doublets~\cite{Atre:2008iu} (the small deviations, with no
consequence at the LHC, are due to the heavy modes with masses $\gtrsim 3$
TeV). 
As we see, all four new leptons decay into taus 100 \% of the time, 
$N\to W^+ \tau$, $Y\to W^- \tau$, $T^1 \to Z \tau$ and $T^2 \to H \tau$. 

The new leptons are produced in pairs at the LHC. 
Single production in association with a $\tau$ is suppressed 
because the off-diagonal couplings $\bar{\tau} FV$, with $F$ the 
new lepton and $V=W, Z$, are small \cite{delAguila:2007em}. 
Drell-Yan pair production results in different final states
\begin{equation}
\tau \bar{\tau} W Z
\mbox{  and  }
\tau \bar{\tau} W H 
\end{equation}
from $W$ exchange, 
\begin{equation}
\tau \bar{\tau} W^+ W^-,~
\tau \bar{\tau} Z Z 
\mbox{  and  }
\tau \bar{\tau} H H 
\end{equation}
from photon and $Z$ exchange, whereas
\begin{equation}
\tau \bar{\tau} Z H,
\end{equation}
only proceeds through $Z$ exchange. 
These signals are difficult to disentangle from the
background because tau leptons must be reconstructed, 
but the relative lightness of these new leptons and their very 
characteristic decay channels help to search for them. 
A dedicated analysis, that is currently under way, is required 
to fully assess the LHC reach for vector-like leptonic resonances 
as predicted in this class of models. (See \cite{delAguila:1989rq} 
for generic 
decay channels.)

\section{Conclusions \label{conclusions}}

Models with warped extra dimensions provide a neat solution to the
hierarchy problem. In this context, models of gauge-Higgs unification
in warped extra dimensions are among the most natural models of EWSB,
realizing in a calculable way the old idea of composite Higgs. 
Furthermore, they also offer a rationale for the flavor structure in
the quark sector. Mass hierarchies and hierarchical mixing angles are
naturally generated by wave function localization, with an added bonus
in the form of a built-in flavor protection that makes new physics at 
$\sim 2-3$ TeV compatible with current EW precision and flavor
data.~\footnote{Full compatibility with flavor data requires a mild
tuning of parameters or some (minimal) structure in the flavor 
realization~\cite{flavor,Csaki:2008zd,Cacciapaglia:2007fw,Santiago:2008vq}.} 
Given the success of GHU models in the quark sector, it
is interesting to study their implications for the leptonic one. 

In this article, we have studied for the first
time the implementation of a global $A_4$ symmetry in models of
GHU. The extra structure implied by the larger gauge symmetry results in some technical differences with respect to the simpler cases 
studied in the past. 
Thus, although LFV is generated at tree level, the global symmetry 
provides a strong enough flavor protection because 
a subgroup of the custodial symmetry required by EWPT 
naturally provides the necessary extra suppression. 
We have also investigated possible deviations from TBM 
(which is predicted at LO by the assumed discrete symmetry breaking) 
and the implications of EWPT, LFV and neutrino masses and mixing 
on the spectrum of new resonances. 
This requires a precise enough determination of the masses 
and mixings of particles spreaded by many orders of magnitude, 
making the numerical analysis rather challenging. 
We must ensure that the many different types 
of corrections do not alter the necessary disparity of masses 
and mixings. 
In particular, the stringent bounds on LFV processes demands 
a precise evaluation of the mixing between light and heavy 
leptons, because this mixing is what renders the one-loop 
contributions small enough.   

The model is compatible with all those experimental constraints 
for new gauge boson masses 
\begin{equation}
M_{KK}^{\mathrm{gauge}}\gtrsim 3.5~\mbox{TeV}. 
\end{equation}
Then, KK gluons could be accessible at the LHC~\cite{Agashe:2006hk}
(see~\cite{Ledroit:2007ik} for KK EW resonances).
A new characteristic feature of our construction is the 
correlation between one-loop LFV (for instance, $\mu \to e \gamma$) 
and the presence of light leptonic resonances in the spectrum. 
In order to keep LFV below current (and expected) experimental bounds, 
the $A_4$ breaking has to be relatively small. 
On the other hand, charged leptons masses are protected by this 
global flavor symmetry. Thus, the smaller its breaking is , 
the more composite $\tau_R$ has to be in order to predict the correct  
$\tau$ mass. This in turn implies the existence of new
leptonic resonances with masses of few hundreds of GeV and large
couplings to $\tau_R$. 
They come in a full almost degenerate $SU(2)_L \times SU(2)_R$ 
bi-doublet with a very distintive phenomenology at the LHC. 
Hence, as the discovery of new resonances in the quark
sector~\cite{Atre:2008iu,Carena:2007tn}, the observation 
of LFV processes near present limits or of 
new vector-like lepton doublets only decaying into taus at LHC 
would be a strong indication of a strongly coupled realization of EWSB. 

\vspace{.4cm}\noindent \textit{Note Added.} 
During the writing of this paper, the possibility of new light
leptonic resonances accessible at LHC 
has been also discussed in models with warped extra
dimensions in~\cite{Agashe:2009ja,Burdman:2009ih}. 
However, in our case these new resonances are a consequence of the $A_4$
symmetry predicting tri-bimaximal mixing, that correlates them to lepton
flavor violating processes through the tau mass.

\begin{acknowledgments}

A.C. thanks ITP at ETH Z\"urich and J.S. thanks GGI 
for hospitality during completion and preliminary presentation of this
work. This work has been
partially supported by MICINN project 
FPA2006-05294, Junta de Andaluc\'{\i}a projects FQM 101 and
FQM03048. The work of J.S. has been partially supported by SNSF
under contract 200021-117873 and by a MICINN \textit{Ram\'on y Cajal}
contract. The work of A.C. has been supported by
a MICINN FPU fellowship.

\end{acknowledgments}

\appendix

\section{Group theory summary}

In this appendix we summarize the main group theory properties used in
the text. 

\subsection{ $A_4$ representations\label{a4}}

$A_4$ is the group of even permutations of four elements. It has twelve 
elements which can be written in terms of two generators, $S$ and $T$,
satisfying 
\begin{eqnarray}
S^2=T^3=(ST)^3=\mathbf{1}.
\end{eqnarray}
This discrete group has three inequivalent one-dimensional representations 
\begin{eqnarray}
\begin{array}{rcl}
\mathbf{1}:&~S=1,&~ T=1,\\
\mathbf{1}^{\prime}:&~S=1,&~ T=e^{i2\pi/3}=\omega,\\
\mathbf{1}^{\prime\prime}:&~S=1,&~ T=e^{i4\pi/3}=\omega^2,
\end{array} 
\end{eqnarray}
and one three-dimensional irreducible representation, {\bf 3}; 
being the Clebsch-Gordan series of their non-trivial products  
\begin{equation}
\begin{array}{c}
\mathbf{1}^{\prime}\times\mathbf{1}^{\prime}
=\mathbf{1}^{\prime\prime}, \quad
\mathbf{1}^{\prime}\times \mathbf{1}^{\prime \prime}=\mathbf{1}, \quad
\mathbf{1}^{\prime\prime}\times\mathbf{1}^{\prime\prime}
=\mathbf{1}^{\prime}, \\ 
\mathbf{1}^{\prime}\times\mathbf{3}=\mathbf{3},  \quad
\mathbf{1}^{\prime\prime}\times \mathbf{3}=\mathbf{3},\\
\mathbf{3}_{x}\times \mathbf{3}_y=
\mathbf{3}_1+\mathbf{3}_2+\mathbf{1}
+\mathbf{1}^{\prime}+\mathbf{1}^{\prime\prime} .
\end{array}
\end{equation}
In the basis where $S$ is diagonal
\begin{eqnarray}
S=\left(\begin{array}{ccc}
 1 & 0 & 0 \\
 0 & -1 & 0 \\
 0 & 0 & -1
\end{array}
\right),\qquad T=\left(
\begin{array}{ccc}
 0 & 1 & 0 \\
 0 & 0 & 1 \\
 1 & 0 & 0
\end{array}
\right),
\end{eqnarray}
and the decomposition of
$\mathbf{3}_x \times \mathbf{3}_y$ reads 
\begin{eqnarray}
\mathbf{1}&=&x_1y_1+x_2y_2+x_3y_3,\nonumber\\
\mathbf{1}^{\prime}&=&x_1y_1+\omega^2 x_2y_2+\omega x_3y_3,\nonumber\\
\mathbf{1}^{\prime \prime}&=&x_1y_1+\omega x_2y_2+\omega^2 x_3y_3,\\
\mathbf{3}_1&=&(x_2y_3,x_3y_1,x_1y_2),\nonumber\\
\mathbf{3}_2&=&(x_3y_2,x_1y_3,x_2y_1),\nonumber
\end{eqnarray}
with  
$\mathbf{3}_x=(x_1,x_2,x_3)$ and
$\mathbf{3}_y=(y_1,y_2,y_3)$.

\subsection{$SO(5)$ generators in the fundamental representation\label{SO5}}

The ten $SO(5)$ generators can be written in the 
fundamental representation (\textbf{5}) 
\begin{eqnarray}
\label{tele}
T_{L,ij}^a&=&-\frac{i}{2}\left[\frac{1}{2}\epsilon^{abc}\left(\delta_i^b
  \delta_j^c-\delta_j^b\delta_i^c\right)
+\left(\delta_i^a\delta_j^4-\delta_j^a\delta_i^4\right)\right],\qquad
a=1,2,3, \nonumber \\ 
\label{tere}
T_{R,ij}^a&=&-\frac{i}{2}\left[\frac{1}{2}\epsilon^{abc}\left(\delta_i^b
  \delta_j^c-\delta_j^b\delta_i^c\right)-\left(\delta_i^a\delta_j^4
-\delta_j^a\delta_i^4\right)\right],\qquad
a=1,2,3, \\ 
T_{C,ij}^{\hat{a}}&=&
-\frac{i}{\sqrt{2}}\left[\delta_i^{\hat{a}}\delta_j^5
-\delta_j^{\hat{a}}\delta_i^5\right],\qquad
\hat{a}=1,2,3,4. \nonumber
\end{eqnarray}
They are normalized to $\textrm{Tr}T^{\alpha}
T^{\beta}=\delta^{\alpha\beta}$. 
In this basis {\bf 5}, which decomposes into 
$(\mathbf{2},\mathbf{2}) \oplus (\mathbf{1},\mathbf{1})$
under $SU(2)_L \times SU(2)_R$, reads
\begin{eqnarray}
Q=\frac{1}{\sqrt{2}}\left(\begin{array}{c}q_{++}-q_{--}\\
iq_{++}+iq_{--}\\-iq_{+-}+iq_{-+}\\q_{+-}+q_{-+}\\\sqrt{2}q_{00}
\end{array}\right),
\end{eqnarray}
where the first (second) subscript $\pm,0$ corresponds to 
$T^3_L = \pm \frac{1}{2},0$  ($T^3_R = \pm \frac{1}{2},0$), respectively.

\section{KK expansion in the presence of boundary 
Majorana masses \label{appendix:majorana}}

Let us consider the effect on the KK expansion of a bulk fermion 
$\psi$ with a UV localized Majorana mass term 
\begin{equation}
\theta_{ij}\overline{\psi}^c_{iR}\psi_{jR}
+\mathrm{h.c.},
\end{equation} 
and then satisfying the UV boundary condition
\begin{eqnarray}
\label{UVbc}
\psi_{iL}(x,R)+\theta_{ij}^\dagger\psi^c_{jR}(x,R)=0.
\end{eqnarray}
The corresponding KK expansion can be written 
\begin{eqnarray}
\psi_{iL}(x,z)=\sum_n g_{in}(z)\xi_n(x) ,
\qquad 
\psi_{iR}(x,z)=\sum_n f_{in}(z)\bar{\xi}_n(x) ,
\end{eqnarray}
where the 4D fields are assumed to obey a Majorana equation 
\begin{eqnarray}
i\bar{\sigma}^{\mu}\partial_{\mu}\xi_n-m_n\bar{\xi}_n=0,
\qquad 
i\sigma^{\mu}\partial_{\mu}\bar{\xi}_n-m_n^{\ast}\xi_n=0.
\end{eqnarray}
If we insert these expansions in the 5D equations of motion and use 
the 4D Majorana equations, we get a system of coupled differential equations 
\begin{eqnarray}
f^{\prime}_{in}+m_ng_{in}-\frac{c_i+2}{z}f_{in}=0,\\
g^{\prime}_{in}-m_n^{\ast}f_{in}+\frac{c_i-2}{z}g_{in}=0.
\end{eqnarray}
The solution of this system of equations can be obtained 
using standard techniques (decoupling by iteration and use of the
first order equation to obtain the second solution), 
\begin{eqnarray}
g_{in}(z)&=&
z^{5/2}\left(A_{in} J_{c_i+1/2}(|m_n|z)
+B_{in} Y_{c_i+1/2}(|m_n| z)\right),\\
f_{in}(z)&=&
\frac{m_n}{|m_n|}z^{5/2}\left(A_{in} J_{c_i-1/2}(|m_n| z)
+ B_{in} Y_{c_i-1/2}(|m_n| z)\right).
\end{eqnarray}
These two $n$ functions are mixed by the boundary condition 
in Eq. (\ref{UVbc}). 
If $\theta_{ij}$ is real, 
$m_n$ will be also real and the linear system resulting from 
imposing the boundary condition will factorize into two simpler ones; 
one for the real part of the unknowns, 
and another one for their imaginary parts 
(obtained changing the sign of $\theta_{ij}$). 




\begin{thebibliography}{99}

\bibitem{Kaplan:1983fs}
  D.~B.~Kaplan and H.~Georgi,
  Phys.\ Lett.\  B {\bf 136} (1984) 183;
  D.~B.~Kaplan, H.~Georgi and S.~Dimopoulos,
  Phys.\ Lett.\  B {\bf 136} (1984) 187.



\bibitem{Maldacena:1997re}
  J.~M.~Maldacena,
  Adv.\ Theor.\ Math.\ Phys.\  {\bf 2} (1998) 231
  [Int.\ J.\ Theor.\ Phys.\  {\bf 38} (1999) 1113]
  [hep-th/9711200];
  S.~S.~Gubser, I.~R.~Klebanov and A.~M.~Polyakov,
  Phys.\ Lett.\  B {\bf 428} (1998) 105
  [hep-th/9802109];
  E.~Witten,
  Adv.\ Theor.\ Math.\ Phys.\  {\bf 2} (1998) 253
  [hep-th/9802150].

\bibitem{Randall:1999vf}
  L.~Randall and R.~Sundrum,
  Phys.\ Rev.\ Lett.\  {\bf 83} (1999) 4690
  [hep-th/9906064];
  Phys.\ Rev.\ Lett.\  {\bf 83} (1999) 3370
  [hep-ph/9905221].


\bibitem{ArkaniHamed:2000ds}
  N.~Arkani-Hamed, M.~Porrati and L.~Randall,
  JHEP {\bf 0108} (2001) 017
  [hep-th/0012148];
  R.~Rattazzi and A.~Zaffaroni,
  JHEP {\bf 0104} (2001) 021
  [hep-th/0012248];
  M.~Perez-Victoria,
  JHEP {\bf 0105} (2001) 064
  [hep-th/0105048].


\bibitem{Contino:2003ve}
  R.~Contino, Y.~Nomura and A.~Pomarol,
  Nucl.\ Phys.\  B {\bf 671} (2003) 148
  [arXiv:hep-ph/0306259].

\bibitem{compositeH:4D}
  G.~F.~Giudice, C.~Grojean, A.~Pomarol and R.~Rattazzi,
  JHEP {\bf 0706} (2007) 045
  [arXiv:hep-ph/0703164];
  R.~Barbieri, B.~Bellazzini, V.~S.~Rychkov and A.~Varagnolo,
  Phys.\ Rev.\  D {\bf 76} (2007) 115008
  [arXiv:0706.0432 [hep-ph]];
  B.~Bellazzini, S.~Pokorski, V.~S.~Rychkov and A.~Varagnolo,
  JHEP {\bf 0811} (2008) 027
  [arXiv:0805.2107 [hep-ph]];
  P.~Lodone,
  JHEP {\bf 0812} (2008) 029
  [arXiv:0806.1472 [hep-ph]];
  A.~Pomarol and J.~Serra,
  Phys.\ Rev.\  D {\bf 78} (2008) 074026
  [arXiv:0806.3247 [hep-ph]];
  M.~Gillioz,
  arXiv:0806.3450 [hep-ph];
  C.~Anastasiou, E.~Furlan and J.~Santiago,
  arXiv:0901.2117 [hep-ph];
  B.~Gripaios, A.~Pomarol, F.~Riva and J.~Serra,
  JHEP {\bf 0904} (2009) 070
  [arXiv:0902.1483 [hep-ph]].




\bibitem{Davoudiasl:2009cd}
  H.~Davoudiasl, S.~Gopalakrishna, E.~Ponton and J.~Santiago,
  arXiv:0908.1968 [hep-ph].




\bibitem{Agashe:2003zs}
  K.~Agashe, A.~Delgado, M.~J.~May and R.~Sundrum,
  JHEP {\bf 0308} (2003) 050
  [arXiv:hep-ph/0308036].



\bibitem{Agashe:2006at}
  K.~Agashe, R.~Contino, L.~Da Rold and A.~Pomarol,
  Phys.\ Lett.\  B {\bf 641} (2006) 62
  [arXiv:hep-ph/0605341].

\bibitem{Djouadi:2006rk}
  A.~Djouadi, G.~Moreau and F.~Richard,
  Nucl.\ Phys.\  B {\bf 773}, 43 (2007)
  [arXiv:hep-ph/0610173].



\bibitem{Agashe:2004rs}
  K.~Agashe, R.~Contino and A.~Pomarol,
  Nucl.\ Phys.\  B {\bf 719} (2005) 165
  [arXiv:hep-ph/0412089].



\bibitem{Medina:2007hz}
  A.~D.~Medina, N.~R.~Shah and C.~E.~M.~Wagner,
  Phys.\ Rev.\  D {\bf 76} (2007) 095010
  [arXiv:0706.1281 [hep-ph]].




\bibitem{Csaki:2008zd}
  C.~Csaki, A.~Falkowski and A.~Weiler,
  JHEP {\bf 0809} (2008) 008
  [arXiv:0804.1954 [hep-ph]].


\bibitem{Agashe:2005dk}
  K.~Agashe and R.~Contino,
  Nucl.\ Phys.\  B {\bf 742} (2006) 59
  [arXiv:hep-ph/0510164].


\bibitem{Carena:2006bn}
  M.~S.~Carena, E.~Ponton, J.~Santiago and C.~E.~M.~Wagner,
  Nucl.\ Phys.\  B {\bf 759} (2006) 202
  [arXiv:hep-ph/0607106];
  Phys.\ Rev.\  D {\bf 76} (2007) 035006
  [arXiv:hep-ph/0701055].


\bibitem{Agashe:2007jb}
  K.~Agashe, A.~Falkowski, I.~Low and G.~Servant,
  JHEP {\bf 0804} (2008) 027
  [arXiv:0712.2455 [hep-ph]].


\bibitem{Panico:2008bx}
  G.~Panico, E.~Ponton, J.~Santiago and M.~Serone,
  Phys.\ Rev.\  D {\bf 77} (2008) 115012
  [arXiv:0801.1645 [hep-ph]].

\bibitem{Carena:2009yt}
  M.~Carena, A.~D.~Medina, N.~R.~Shah and C.~E.~M.~Wagner,
  arXiv:0901.0609 [hep-ph].


\bibitem{Grossman:1999ra}
  Y.~Grossman and M.~Neubert,
  Phys.\ Lett.\  B {\bf 474} (2000) 361
  [arXiv:hep-ph/9912408].


\bibitem{Huber:2000ie}
  S.~J.~Huber and Q.~Shafi,
  Phys.\ Lett.\  B {\bf 498} (2001) 256
  [arXiv:hep-ph/0010195];
  Phys.\ Lett.\  B {\bf 512} (2001) 365
  [arXiv:hep-ph/0104293];
  Phys.\ Lett.\  B {\bf 544} (2002) 295
  [arXiv:hep-ph/0205327].

\bibitem{Huber:2003sf}
  S.~J.~Huber and Q.~Shafi,
  Phys.\ Lett.\  B {\bf 583} (2004) 293
  [arXiv:hep-ph/0309252].

\bibitem{ArkaniHamed:1999dc}
  N.~Arkani-Hamed and M.~Schmaltz,
  Phys.\ Rev.\  D {\bf 61} (2000) 033005
  [arXiv:hep-ph/9903417].

\bibitem{Agashe:2008fe}
  K.~Agashe, T.~Okui and R.~Sundrum,
  arXiv:0810.1277 [hep-ph].


\bibitem{A4in4D}
  E.~Ma and G.~Rajasekaran,
  Phys.\ Rev.\  D {\bf 64}, 113012 (2001)
  [arXiv:hep-ph/0106291];
  K.~S.~Babu, E.~Ma and J.~W.~F.~Valle,
  Phys.\ Lett.\  B {\bf 552}, 207 (2003)
  [arXiv:hep-ph/0206292];
  M.~Hirsch, J.~C.~Romao, S.~Skadhauge, J.~W.~F.~Valle and
  A.~Villanova del Moral, 
  Phys.\ Rev.\  D {\bf 69}, 093006 (2004)
  [arXiv:hep-ph/0312265];
  E.~Ma,
  Phys.\ Rev.\  D {\bf 70}, 031901 (2004)
  [arXiv:hep-ph/0404199];
  S.~L.~Chen, M.~Frigerio and E.~Ma,
  Nucl.\ Phys.\  B {\bf 724}, 423 (2005)
  [arXiv:hep-ph/0504181];
  E.~Ma,
  Phys.\ Rev.\  D {\bf 73}, 057304 (2006)
  [arXiv:hep-ph/0511133];
  G.~Altarelli and F.~Feruglio,
  Nucl.\ Phys.\  B {\bf 741}, 215 (2006)
  [arXiv:hep-ph/0512103];
 I.~de Medeiros Varzielas, S.~F.~King and G.~G.~Ross,
 Phys.\ Lett.\  B {\bf 644}, 153 (2007)
 [arXiv:hep-ph/0512313];
  X.~G.~F.~He, Y.~Y.~F.~Keum and R.~R.~Volkas,
  JHEP {\bf 0604}, 039 (2006)
  [arXiv:hep-ph/0601001];
  B.~Adhikary, B.~Brahmachari, A.~Ghosal, E.~Ma and M.~K.~Parida,
  Phys.\ Lett.\  B {\bf 638}, 345 (2006)
  [arXiv:hep-ph/0603059];
  L.~Lavoura and H.~Kuhbock,
  Mod.\ Phys.\ Lett.\  A {\bf 22}, 181 (2007)
  [arXiv:hep-ph/0610050];
  S.~F.~King and M.~Malinsky,
  Phys.\ Lett.\  B {\bf 645}, 351 (2007)
  [arXiv:hep-ph/0610250];
  S.~Morisi, M.~Picariello and E.~Torrente-Lujan,
  Phys.\ Rev.\  D {\bf 75}, 075015 (2007)
  [arXiv:hep-ph/0702034];
  F.~Yin,
  Phys.\ Rev.\  D {\bf 75}, 073010 (2007)
  [arXiv:0704.3827 [hep-ph]];
  F.~Bazzocchi, S.~Kaneko and S.~Morisi,
  JHEP {\bf 0803}, 063 (2008)
  [arXiv:0707.3032 [hep-ph]];
  W.~Grimus and H.~Kuhbock,
  Phys.\ Rev.\  D {\bf 77}, 055008 (2008)
  [arXiv:0710.1585 [hep-ph]];
  F.~Bazzocchi, S.~Morisi and M.~Picariello,
  Phys.\ Lett.\  B {\bf 659}, 628 (2008)
  [arXiv:0710.2928 [hep-ph]];
  M.~Honda and M.~Tanimoto,
  Prog.\ Theor.\ Phys.\  {\bf 119}, 583 (2008)
  [arXiv:0801.0181 [hep-ph]];
  B.~Brahmachari, S.~Choubey and M.~Mitra,
  Phys.\ Rev.\  D {\bf 77}, 073008 (2008)
  [Erratum-ibid.\  D {\bf 77}, 119901 (2008)]
  [arXiv:0801.3554 [hep-ph]];
  G.~Altarelli, F.~Feruglio and C.~Hagedorn,
  JHEP {\bf 0803}, 052 (2008)
  [arXiv:0802.0090 [hep-ph]];
  F.~Bazzocchi, S.~Morisi, M.~Picariello and E.~Torrente-Lujan,
  J.\ Phys.\ G {\bf 36}, 015002 (2009)
  [arXiv:0802.1693 [hep-ph]];
  B.~Adhikary and A.~Ghosal,
  Phys.\ Rev.\  D {\bf 78}, 073007 (2008)
  [arXiv:0803.3582 [hep-ph]];
  Y.~Lin,
  Nucl.\ Phys.\  B {\bf 813}, 91 (2009)
  [arXiv:0804.2867 [hep-ph]];
  F.~Bazzocchi, M.~Frigerio and S.~Morisi,
  Phys.\ Rev.\  D {\bf 78}, 116018 (2008)
  [arXiv:0809.3573 [hep-ph]];
  S.~Morisi,
  Phys.\ Rev.\  D {\bf 79}, 033008 (2009)
  [arXiv:0901.1080 [hep-ph]];
  P.~Ciafaloni, M.~Picariello, E.~Torrente-Lujan and A.~Urbano,
  Phys.\ Rev.\  D {\bf 79}, 116010 (2009)
  [arXiv:0901.2236 [hep-ph]];
  M.~C.~Chen and S.~F.~King,
  JHEP {\bf 0906}, 072 (2009)
  [arXiv:0903.0125 [hep-ph]];
  G.~C.~Branco, R.~Gonzalez Felipe, M.~N.~Rebelo and H.~Serodio,
  Phys.\ Rev.\  D {\bf 79}, 093008 (2009)
  [arXiv:0904.3076 [hep-ph]];
  A.~Hayakawa, H.~Ishimori, Y.~Shimizu and M.~Tanimoto,
  Phys.\ Lett.\  B {\bf 680}, 334 (2009)
  [arXiv:0904.3820 [hep-ph]];
  G.~Altarelli and D.~Meloni,
  J.\ Phys.\ G {\bf 36}, 085005 (2009)
  [arXiv:0905.0620 [hep-ph]];
  Y.~Lin,
  Nucl.\ Phys.\  B {\bf 824}, 95 (2010)
  [arXiv:0905.3534 [hep-ph]];
  C.~Hagedorn, E.~Molinaro and S.~T.~Petcov,
  JHEP {\bf 0909}, 115 (2009)
  [arXiv:0908.0240 [hep-ph]];
  E.~Ma,
  arXiv:0908.3165 [hep-ph];
  T.~J.~Burrows and S.~F.~King,
  arXiv:0909.1433 [hep-ph];
  A.~Albaid,
  Phys.\ Rev.\  D {\bf 80}, 093002 (2009)
  [arXiv:0909.1762 [hep-ph]];
  P.~Ciafaloni, M.~Picariello, E.~Torrente-Lujan and A.~Urbano,
  arXiv:0909.2553 [hep-ph];
  F.~Feruglio, C.~Hagedorn and L.~Merlo,
  arXiv:0910.4058;
  S.~Morisi and E.~Peinado,
  arXiv:0910.4389;
  J.~Berger and Y.~Grossman,
  arXiv:0910.4392 [hep-ph].




\bibitem{LFV:A4}
  F.~Feruglio, C.~Hagedorn, Y.~Lin and L.~Merlo,
  Nucl.\ Phys.\  B {\bf 809}, 218 (2009)
  [arXiv:0807.3160 [hep-ph]].
  C.~Hagedorn, E.~Molinaro and S.~T.~Petcov,
  arXiv:0911.3605;
  F.~Feruglio, C.~Hagedorn, Y.~Lin and L.~Merlo,
  arXiv:0911.3874;
  G.~J.~Ding and J.~F.~Liu,
  arXiv:0911.4799.


\bibitem{Harrison:2002er}
  P.~F.~Harrison, D.~H.~Perkins and W.~G.~Scott,
  Phys.\ Lett.\  B {\bf 530}, 167 (2002)
  [arXiv:hep-ph/0202074].



\bibitem{GonzalezGarcia:2007ib}
  M.~C.~Gonzalez-Garcia and M.~Maltoni,
  Phys.\ Rept.\  {\bf 460}, 1 (2008)
  [arXiv:0704.1800 [hep-ph]].

\bibitem{Schwetz:2008er}
  T.~Schwetz, M.~A.~Tortola and J.~W.~F.~Valle,
  New J.\ Phys.\  {\bf 10}, 113011 (2008)
  [arXiv:0808.2016 [hep-ph]];
  G.~L.~Fogli, E.~Lisi, A.~Marrone, A.~Palazzo and A.~M.~Rotunno,
  arXiv:0809.2936 [hep-ph].




\bibitem{Csaki:2008qq}
  C.~Csaki, C.~Delaunay, C.~Grojean and Y.~Grossman,
  JHEP {\bf 0810} (2008) 055
  [arXiv:0806.0356 [hep-ph]].


\bibitem{Feruglio:2007uu}
  F.~Feruglio, C.~Hagedorn, Y.~Lin and L.~Merlo,
  Nucl.\ Phys.\  B {\bf 775}, 120 (2007)
  [arXiv:hep-ph/0702194].

\bibitem{Chen:2009gy}
  M.~C.~Chen, K.~T.~Mahanthappa and F.~Yu,
  arXiv:0907.3963 [hep-ph];
  arXiv:0909.5472 [hep-ph].

\bibitem{deMedeirosVarzielas:2005ax}
 I.~de Medeiros Varzielas and G.~G.~Ross,
 Nucl.\ Phys.\  B {\bf 733}, 31 (2006)
 [arXiv:hep-ph/0507176];
 I.~de Medeiros Varzielas, S.~F.~King and G.~G.~Ross,
 Phys.\ Lett.\  B {\bf 648}, 201 (2007)
 [arXiv:hep-ph/0607045].


\bibitem{Atre:2008iu}
  A.~Atre, M.~Carena, T.~Han and J.~Santiago,
  Phys.\ Rev.\  D {\bf 79}, 054018 (2009)
  [arXiv:0806.3966 [hep-ph]].


\bibitem{Agashe:2009tu}
  K.~Agashe,
  arXiv:0902.2400 [hep-ph].

\bibitem{Buras:2009ka}
  A.~J.~Buras, B.~Duling and S.~Gori,
  JHEP {\bf 0909}, 076 (2009)
  [arXiv:0905.2318 [hep-ph]].


\bibitem{vonGersdorff:2002rg}
  G.~von Gersdorff, N.~Irges and M.~Quiros,
  arXiv:hep-ph/0206029.


\bibitem{Contino:2006qr}
  R.~Contino, L.~Da Rold and A.~Pomarol,
  Phys.\ Rev.\  D {\bf 75} (2007) 055014
  [arXiv:hep-ph/0612048].


\bibitem{Santiago:2008vq}
  J.~Santiago,
  JHEP {\bf 0812}, 046 (2008)
  [arXiv:0806.1230 [hep-ph]].


\bibitem{delAguila:2000kb}
  F.~del Aguila and J.~Santiago,
  Phys.\ Lett.\  B {\bf 493}, 175 (2000)
  [arXiv:hep-ph/0008143];
  arXiv:hep-ph/0011143.


\bibitem{Georgi:2000ks}
  H.~Georgi, A.~K.~Grant and G.~Hailu,
  Phys.\ Lett.\  B {\bf 506}, 207 (2001)
  [arXiv:hep-ph/0012379];
  M.~S.~Carena, T.~M.~P.~Tait and C.~E.~M.~Wagner,
  Acta Phys.\ Polon.\  B {\bf 33}, 2355 (2002)
  [arXiv:hep-ph/0207056].




\bibitem{BKT}
  H.~Davoudiasl, J.~L.~Hewett and T.~G.~Rizzo,
  Phys.\ Rev.\  D {\bf 68}, 045002 (2003)
  [arXiv:hep-ph/0212279];
  M.~S.~Carena, E.~Ponton, T.~M.~P.~Tait and C.~E.~M.~Wagner,
  Phys.\ Rev.\  D {\bf 67}, 096006 (2003)
  [arXiv:hep-ph/0212307];
  F.~del Aguila, M.~Perez-Victoria and J.~Santiago,
  JHEP {\bf 0302}, 051 (2003)
  [arXiv:hep-th/0302023];
  Acta Phys.\ Polon.\  B {\bf 34}, 5511 (2003)
  [arXiv:hep-ph/0310353];
  JHEP {\bf 0610}, 056 (2006)
  [arXiv:hep-ph/0601222];
  M.~S.~Carena, A.~Delgado, E.~Ponton, T.~M.~P.~Tait and C.~E.~M.~Wagner,
  Phys.\ Rev.\  D {\bf 68}, 035010 (2003)
  [arXiv:hep-ph/0305188];
  Phys.\ Rev.\  D {\bf 71}, 015010 (2005)
  [arXiv:hep-ph/0410344].


\bibitem{Agashe:2006iy}
  K.~Agashe, A.~E.~Blechman and F.~Petriello,
  Phys.\ Rev.\  D {\bf 74}, 053011 (2006)
  [arXiv:hep-ph/0606021].

\bibitem{Tommasini:1995ii}
  D.~Tommasini, G.~Barenboim, J.~Bernabeu and C.~Jarlskog,
  Nucl.\ Phys.\  B {\bf 444}, 451 (1995)
  [arXiv:hep-ph/9503228];
  J.~I.~Illana and T.~Riemann,
  Phys.\ Rev.\  D {\bf 63}, 053004 (2001)
  [arXiv:hep-ph/0010193].

\bibitem{Appelquist:1974tg}
  T.~Appelquist and J.~Carazzone,
  Phys.\ Rev.\  D {\bf 11}, 2856 (1975).

\bibitem{Amsler:2008zzb}
  C.~Amsler {\it et al.}  [Particle Data Group],
  Phys.\ Lett.\  B {\bf 667}, 1 (2008).


\bibitem{delAguila:2008pw}
  F.~del Aguila, J.~de Blas and M.~Perez-Victoria,
  Phys.\ Rev.\  D {\bf 78}, 013010 (2008)
  [arXiv:0803.4008 [hep-ph]];
  F.~del Aguila, J.~A.~Aguilar-Saavedra and J.~de Blas,
  Acta Phys.\ Polon.\  B {\bf 40}, 2901 (2009)
  [arXiv:0910.2720 [hep-ph]].



\bibitem{Agashe:2009di}
  K.~Agashe and R.~Contino,
  Phys.\ Rev.\  D {\bf 80}, 075016 (2009)
  [arXiv:0906.1542 [hep-ph]];
  A.~Azatov, M.~Toharia and L.~Zhu,
  Phys.\ Rev.\  D {\bf 80}, 035016 (2009)
  [arXiv:0906.1990 [hep-ph]].

\bibitem{Kuno:1999jp}
  Y.~Kuno and Y.~Okada,
  Rev.\ Mod.\ Phys.\  {\bf 73} (2001) 151
  [arXiv:hep-ph/9909265].


\bibitem{Wintz:1998rp}
  P.~Wintz,
{\it Prepared for 29th International Conference on High-Energy Physics
  (ICHEP 98), Vancouver, Canada, 23-29 Jul 1998} 

\bibitem{delAguila:1982yu}
  F.~del Aguila and M.~J.~Bowick,
  Phys.\ Lett.\  B {\bf 119}, 144 (1982).


\bibitem{Blanke:2007db}
  M.~Blanke, A.~J.~Buras, B.~Duling, A.~Poschenrieder and C.~Tarantino,
  JHEP {\bf 0705}, 013 (2007)
  [arXiv:hep-ph/0702136];
  T.~Goto, Y.~Okada and Y.~Yamamoto,
  Phys.\ Lett.\  B {\bf 670}, 378 (2009)
  [arXiv:0809.4753 [hep-ph]];
  F.~del Aguila, J.~I.~Illana and M.~D.~Jenkins,
  JHEP {\bf 0901}, 080 (2009)
  [arXiv:0811.2891 [hep-ph]].


\bibitem{delAguila:1982fs}
  F.~del Aguila and M.~J.~Bowick,
  Nucl.\ Phys.\  B {\bf 224}, 107 (1983);
  F.~del Aguila, M.~Perez-Victoria and J.~Santiago,
  JHEP {\bf 0009}, 011 (2000)
  [arXiv:hep-ph/0007316];



\bibitem{Maki:2008zz}
  A.~Maki,
  AIP Conf.\ Proc.\  {\bf 981} (2008) 363.





\bibitem{DelAguila:2001pu}
  F.~Del Aguila and J.~Santiago,
  JHEP {\bf 0203}, 010 (2002)
  [arXiv:hep-ph/0111047];
  K.~Agashe and G.~Servant,
  JCAP {\bf 0502}, 002 (2005)
  [arXiv:hep-ph/0411254].


\bibitem{delAguila:2007em}
  F.~del Aguila, J.~A.~Aguilar-Saavedra and R.~Pittau,
  JHEP {\bf 0710}, 047 (2007)
  [arXiv:hep-ph/0703261];
  F.~del Aguila and J.~A.~Aguilar-Saavedra,
  Nucl.\ Phys.\  B {\bf 813}, 22 (2009)
  [arXiv:0808.2468 [hep-ph]].

\bibitem{delAguila:1989rq}
  F.~del Aguila, L.~Ametller, G.~L.~Kane and J.~Vidal,
  Nucl.\ Phys.\  B {\bf 334}, 1 (1990);
  J.~A.~Aguilar-Saavedra,
  Nucl.\ Phys.\  B {\bf 828}, 289 (2010)
  [arXiv:0905.2221 [hep-ph]].


\bibitem{flavor}
  K.~Agashe, G.~Perez and A.~Soni,
  Phys.\ Rev.\  D {\bf 71}, 016002 (2005)
  [arXiv:hep-ph/0408134];
  G.~Moreau and J.~I.~Silva-Marcos,
  JHEP {\bf 0603}, 090 (2006)
  [arXiv:hep-ph/0602155];
  S.~Casagrande, F.~Goertz, U.~Haisch, M.~Neubert and T.~Pfoh,
  JHEP {\bf 0810}, 094 (2008)
  [arXiv:0807.4937 [hep-ph]];
  M.~Blanke, A.~J.~Buras, B.~Duling, S.~Gori and A.~Weiler,
  JHEP {\bf 0903}, 001 (2009)
  [arXiv:0809.1073 [hep-ph]];
  K.~Agashe, A.~Azatov and L.~Zhu,
  Phys.\ Rev.\  D {\bf 79}, 056006 (2009)
  [arXiv:0810.1016 [hep-ph]];
  M.~E.~Albrecht, M.~Blanke, A.~J.~Buras, B.~Duling and K.~Gemmler,
  JHEP {\bf 0909}, 064 (2009)
  [arXiv:0903.2415 [hep-ph]];
  M.~Bauer, S.~Casagrande, U.~Haisch and M.~Neubert,
  arXiv:0912.1625 [hep-ph].

\bibitem{Cacciapaglia:2007fw}
  G.~Cacciapaglia, C.~Csaki, J.~Galloway, G.~Marandella, J.~Terning
  and A.~Weiler, 
  JHEP {\bf 0804} (2008) 006
  [arXiv:0709.1714 [hep-ph]];
  A.~L.~Fitzpatrick, G.~Perez and L.~Randall,
  arXiv:0710.1869 [hep-ph];
  C.~Cheung, A.~L.~Fitzpatrick and L.~Randall,
  JHEP {\bf 0801} (2008) 069
  [arXiv:0711.4421 [hep-th]];
  M.~C.~Chen and H.~B.~Yu,
  arXiv:0804.2503 [hep-ph];
  C.~Csaki, A.~Falkowski and A.~Weiler,
  Phys.\ Rev.\  D {\bf 80}, 016001 (2009)
  [arXiv:0806.3757 [hep-ph]];
  C.~Csaki, G.~Perez, Z.~Surujon and A.~Weiler,
  arXiv:0907.0474 [hep-ph].


\bibitem{Agashe:2006hk}
  K.~Agashe, A.~Belyaev, T.~Krupovnickas, G.~Perez and J.~Virzi,
  Phys.\ Rev.\  D {\bf 77}, 015003 (2008)
  [arXiv:hep-ph/0612015];
  B.~Lillie, L.~Randall and L.~T.~Wang,
  JHEP {\bf 0709}, 074 (2007)
  [arXiv:hep-ph/0701166];
  B.~Lillie, J.~Shu and T.~M.~P.~Tait,
  Phys.\ Rev.\  D {\bf 76}, 115016 (2007)
  [arXiv:0706.3960 [hep-ph]].


\bibitem{Ledroit:2007ik}
  F.~Ledroit, G.~Moreau and J.~Morel,
  JHEP {\bf 0709}, 071 (2007)
  [arXiv:hep-ph/0703262];
  A.~Djouadi, G.~Moreau and R.~K.~Singh,
  Nucl.\ Phys.\  B {\bf 797}, 1 (2008)
  [arXiv:0706.4191 [hep-ph]];
  K.~Agashe {\it et al.},
  Phys.\ Rev.\  D {\bf 76}, 115015 (2007)
  [arXiv:0709.0007 [hep-ph]];
  K.~Agashe, S.~Gopalakrishna, T.~Han, G.~Y.~Huang and A.~Soni,
  arXiv:0810.1497 [hep-ph];
  K.~Agashe, A.~Azatov, T.~Han, Y.~Li, Z.~G.~Si and L.~Zhu,
  arXiv:0911.0059 [hep-ph].

\bibitem{Carena:2007tn}
  M.~Carena, A.~D.~Medina, B.~Panes, N.~R.~Shah and C.~E.~M.~Wagner,
  Phys.\ Rev.\  D {\bf 77}, 076003 (2008)
  [arXiv:0712.0095 [hep-ph]];
  R.~Contino and G.~Servant,
  JHEP {\bf 0806}, 026 (2008)
  [arXiv:0801.1679 [hep-ph]];
  J.~A.~Aguilar-Saavedra,
  arXiv:0907.3155 [hep-ph];
  J.~Mrazek and A.~Wulzer,
  arXiv:0909.3977 [hep-ph].

\bibitem{Agashe:2009ja}
  K.~Agashe, K.~Blum, S.~J.~Lee and G.~Perez,
  arXiv:0912.3070 [hep-ph].

\bibitem{Burdman:2009ih}
  G.~Burdman, L.~Da Rold and R.~D.~Matheus,
  arXiv:0912.5219 [hep-ph].




\end{thebibliography}
\end{document}